\documentclass[aps,prd,superscriptaddress,nofootinbib,twocolumn,10pt]{revtex4-2}
\usepackage{color,amsmath,amssymb,graphicx,latexsym,lineno}
\usepackage{threeparttable,multirow,txfonts,subfigure,booktabs}

\begin{document}

\title{Simulation study of performance of the Very Large Area gamma-ray Space Telescope}

\author{Xu Pan}
\affiliation{Key Laboratory of Dark Matter and Space Astronomy, Purple Mountain Observatory, Chinese Academy of Sciences, Nanjing 210023, P. R. China}
\affiliation{School of Astronomy and Space Science, University of Science and Technology of China, Hefei 230026, P. R. China}

\author{Wei Jiang}
\affiliation{Key Laboratory of Dark Matter and Space Astronomy, Purple Mountain Observatory, Chinese Academy of Sciences, Nanjing 210023, P. R. China}

\author{Chuan Yue}
\affiliation{Key Laboratory of Dark Matter and Space Astronomy, Purple Mountain Observatory, Chinese Academy of Sciences, Nanjing 210023, P. R. China}

\author{Shi-Jun Lei}
\affiliation{Key Laboratory of Dark Matter and Space Astronomy, Purple Mountain Observatory, Chinese Academy of Sciences, Nanjing 210023, P. R. China}

\author{Yu-Xin Cui}
\affiliation{Key Laboratory of Dark Matter and Space Astronomy, Purple Mountain Observatory, Chinese Academy of Sciences, Nanjing 210023, P. R. China}
\affiliation{School of Astronomy and Space Science, University of Science and Technology of China, Hefei 230026, P. R. China}

\author{Qiang  Yuan}
\email[Corresponding author: ]{yuanq@pmo.ac.cn}
\affiliation{Key Laboratory of Dark Matter and Space Astronomy, Purple Mountain Observatory, Chinese Academy of Sciences, Nanjing 210023, P. R. China}
\affiliation{School of Astronomy and Space Science, University of Science and Technology of China, Hefei 230026, P. R. China}

\begin{abstract}
The Very Large Area gamma-ray Space Telescope (VLAST) is a mission concept proposed to detect gamma-ray photons through both the Compton scattering and electron-positron pair production mechanisms, enabling the detection of photons with energies ranging from MeV to TeV. This project aims to conduct a comprehensive survey of the gamma-ray sky from a low Earth orbit using an anti-coincidence detector, a tracker detector that also serves as a low energy calorimeter, and a high energy imaging calorimeter. We developed a Monte Carlo simulation application of the detector with the GEANT4 toolkit to evaluate the instrument performance including the effective area, angular resolution and energy resolution, as well as explored specific optimizations of the detector configuration. Our simulation-based analysis indicates that the VLAST's current design is physically feasible, with an acceptance larger than 10~$\rm m^2\ sr$ which is four times larger than Fermi-LAT, an energy resolution better than 2\% at 10~GeV, and an angular resolution better than 0.2 degrees at 10~GeV. The VLAST project is expected to make significant contribution to the field of gamma-ray astronomy and to enhance our understanding of the cosmos.
\end{abstract}


\maketitle

\section{Introduction}\label{sec.I}

Gamma-ray astrophysics is an exciting field of astronomical sciences that has received a strong impulse. Detecting cosmic gamma-ray emission in the energy range from MeV to GeV can hardly be done by groundbased telescopes, and is preferably to be done in space. Gamma-ray telescopes covering such an energy range can be roughly divided into two categories based on the detection principle: pair production telescopes and Compton scattering telescopes. For the pair production telescopes,  OSO-3 \cite{1972ApJ...177..341K}  provided the first confirmation that the detection of gamma rays is feasible in a complex background of charged particles. The breakthrough discoveries  of high-energy gamma-ray observations were carried out by the SAS-2 \cite{1975ApJ...198..163F} and COS-B \cite{1975SSI.....1..245B} missions in the 1970s. In the 1990s, EGRET made significant progress in surveying the gamma-ray sky above 50~MeV, leading to the discovery of numerous high-energy gamma-ray sources \cite{1993ApJS...86..629T}. The Fermi-LAT gamma-ray space telescope~\cite{2009ApJ...697.1071A}, launched in 2008, has been highly successful in this field for over a decade, identifying more than 6000 gamma-ray sources in its fourth catalog \cite{2022ApJS..260...53A}. However, due to the detector's limited  acceptance and angular resolution, nearly a third of the sources remain unidentified. The GAMMA-400 space mission, set to be installed on the Russian space platform Navigator, is currently under preparation. With its excellent energy resolution and unprecedented angular resolution above 30~GeV compared to other space-based, it has the potential to unlock new insights in this field. However, its effective area is limited to 4000 $\rm cm^2$ due to the detector's size \cite{2022AdSpR..70.2773T}. For the Compton scattering telescope, the pioneering telescope to open the MeV gamma-ray astronomical window was COMPTEL \cite{1993ApJS...86..657S}. Another small detector on board the Chinese space station, POLAR \cite{2020A&A...644A.124K}, is dedicated to measurements of polarization of MeV gamma rays through Compton scattering. The COSI \cite{2021arXiv210213158Z} project, which is funded by NASA's Small Explorer program, is scheduled for launch in 2026 and features exceptional energy resolution. The effective area of COSI is still small, and perhaps does not match the requirement of a powerful detector for MeV time-domain astronomy. Recent advancements in detection technology (semiconductor, scintillator, and time projection chamber) have sparked growing interests in the MeV energy band. As a result, several space-based gamma-ray missions have been proposed in recent years, such as PANGU \cite{2014SPIE.9144E..0FW}, AMEGO \cite{2019BAAS...51g.245M}, e-ASTROGAM \cite{2017ExA....44...25D}, AdEPT \cite{2014APh....59...18H}, GECCO \cite{2022JCAP...07..036O}, MAST \cite{2019APh...112....1D}, GRAMS \cite{2020APh...114..107A}, XGIS-THESEUS \cite{2021arXiv210208701L}, Crystal Eye \cite{2022icrc.confE.581B}, and MASS \cite{2023arXiv231211900Z}. With ongoing developments of detection technology and increasing scientific demands, there is a pressing need for gamma-ray telescopes with enhanced sensitivity.


We propose the Very Large Area gamma-ray Space Telescope (VLAST) \cite{2022AcASn..63...27F,2022cosp...44.3058W,wan2023design,Yang:2022eaq} with a significantly larger effective area which is four times larger than Fermi-LAT. VLAST mainly consists of an Anti-Coincidence Detector (ACD), a Silicon Tracker and low Energy gamma-ray Detector (STED), and an High Energy Imaging  Calorimeter (HEIC), combining the advantages of DAMPE \cite{2017APh....95....6C} and APT \cite{2022icrc.confE.655B} in the design. The STED design can measure both low-energy Compton  scattering and high-energy pair production events, which changed the traditional tungsten plate to thin Cesium Iodide (CsI) tile allowing model-independent control on the detector systematic uncertainties. VLAST can survey the gamma-ray sky from a low Earth orbit in the energy band from about 0.1~MeV to more than 1~TeV.

The key scientific goals of VLAST include (i) searching for the dark matter signatures in the galaxy cores, galaxy disks, and dwarf galaxies \cite{2018PhRvD..97f3003X,2016PhRvD..93j3525L,2012JCAP...07..054B,2013PhRvD..87a5015W,2015PhRvD..91l3010Z}, (ii) monitoring the special gamma-ray sources over time, such as Active Galactic Nuclei, gamma-ray bursts, millisecond pulsars, supernovas, and so on \cite{2010ApJ...724.1044S,2019Natur.575..448Z,2016ApJ...831..143X,2014JHEAp...3....1Y,2016ApJ...823...44X,2015ApJ...809..102L}, (iii) understanding the  origin and transportation of the cosmic rays \cite{2016ApJS..224....8A,2013Sci...339..807A,2017PhRvD..95h3007Y}, (iv) using  extra-galactic diffuse gamma-rays and the gamma-ray horizons to study the evolution of the universe \cite{2022PhRvD.105f3008C,2021Natur.597..341R,2018Sci...362.1031F,2019ApJ...882...87Z}, (v) testing  fundamental physical laws, such as the Lorentz invariance and the equivalence principle \cite{2021FrPhy..1644300W,1998Natur.393..763A,2023APh...14802831L,2022PhRvL.128e1102C}, and so on.
 

In this work, we evaluate the performance and optimize the design of VLAST by simulation. The simulation framework  is developed based on GEANT4 \cite{2016NIMPA.835..186A}. For the event reconstruction, different  energy or trajectories  reconstruction algorithms are used to analyze Compton scattering and pair production events. The key performance parameters of VLAST were obtained based on the detailed MC simulations, such as acceptance, effective area, angular resolution, energy resolution, and e-p discrimination. Furthermore, Some specific design parameters are optimized, such as the threshold of ACD, the size of ACD, the width of the silicon strip pitch, the number of CsI layers in the STED, and so on.

\section{BASELINE DESIGN OF VLAST}\label{sec.II}
\begin{figure*}[!htb]
\centering
\includegraphics[width=1.0\textwidth]{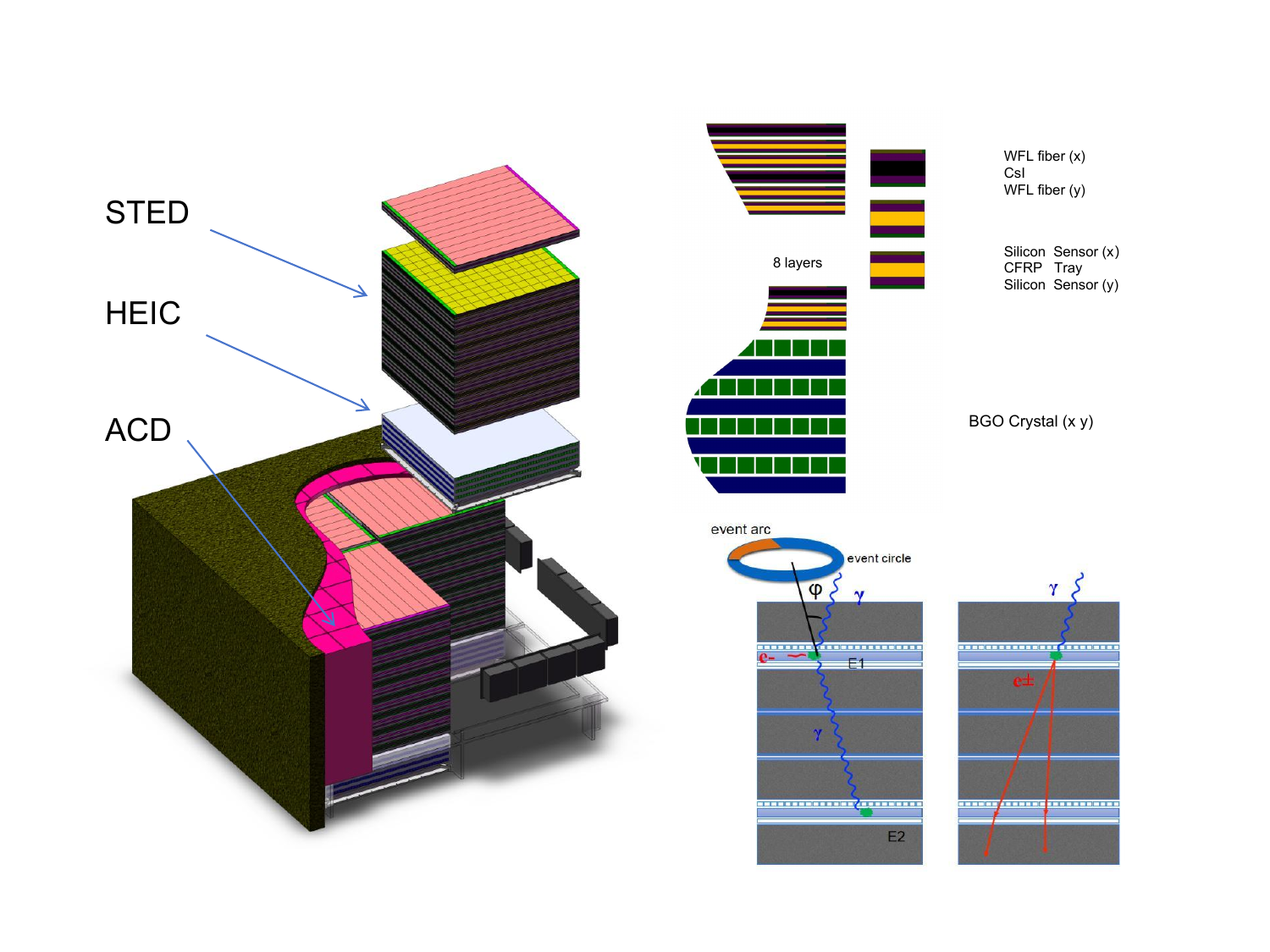}
\caption{The schematic of the VLAST detector. The ACD covers the whole detector except the bottom. The STED has a total of 8 layers, each containing one sub-layer of CsI crystals connected to wavelength-shifting fibers and two sub-layers of silicon strip modules. The HEIC is placed below the STED which contains four layers of orthogonally arranged Bismuth Germanium Oxide (BGO) crystals. The bottom right panel shows the two classes of events detectable by VLAST, the Compton scattering events and the electron-positron pair production events.}
\label{fig:geomtry}
\end{figure*}

VLAST follows the foundational principles and structure of earlier gamma detectors. The composition of the detector includes three primary sub-detectors: ACD, STED, and HEIC. The ACD plays a critical role in rejecting charged particle background and reducing backsplashes from high-energy events. The STED, sharing design similarities with APT, substitutes the tungsten plate in the tracker detector with thin CsI tiles. This alteration allows for the measurement of Compton events and pair production events, enabling the tracking of their trajectories to reconstruct the direction of incident gamma-rays. Low-energy MeV photons are measured primarily through Compton scattering with electrons to determine their energy and direction of incidence. The HEIC measures the energy of incident particles, images the profile of the electromagnetic or hadron shower of high-energy particles, which is used to discriminate between electrons and protons, and provides an estimated direction of the incident particle. Fig.~\ref{fig:geomtry} shows the schematic of  VLAST detector.

\subsection{Anti-Coincidence Detector}
The primary function of the ACD is to minimize the backsplash effect caused by high-energy photons and to reject the background of charged particles. Serving as the first barrier to reject charged particles, it is closely wrapped around the STED. The ACD is composed of 448 plastic scintillator detector tiles (20~cm$\times$20~cm$\times$1~cm). Plastic scintillators have a high detection efficiency for charged particles and a low detection efficiency for photons. So they are often used as ACD for gamma-ray detection. In order to prevent particles from passing through the gap between ACD without being identified, the gaps between tiles are covered by flexible scintillating fiber ribbons.

 The backsplash effect was initially discovered by the EGRET team. When a high-energy photon enters the calorimeter, it generates numerous secondary low-energy photons (ranging from 0.1 to a few MeV), some of which may travel in the opposing direction through the STED and reach the ACD.  This scattering causes the electrons to recoil, generating an electrical signal in the ACD. The intensity of this effect hinges on many factors such as the incident particle's energy, the calorimeter's thickness, and its material composition. The intensity of the backsplash effect increases with the energy of the incident particles and  radiation length of the calorimeter. To mitigate this impact, the Fermi-LAT detector divides the ACD into smaller units, utilizing the discrepancy between the piece in the incident direction and the piece in the recoil to discriminate the recoil event. This modular design effectively suppresses misinterpretations of photons attributed to the backsplash effect of high-energy photons. Following this scheme, the ACD in the VLAST design enables the dismissal of fired tiles not aligned with the direction of photon incidence, thereby significantly mitigating the influence of the backsplash effect. The backsplash effect in VLAST is depicted in Fig.~\ref{fig:backsplash}.

  The photon flux is lower than that of protons by a factor of $10^5$ and less than electrons by about $10^{3}$ in the observational energy band of VLAST. In order to detect photons in the complex cosmic ray background of charged particles, VLAST must have outstanding ability to distinguish photons and charged particles. When photons transform into electron-positron pairs in STED, they can be discriminated from protons based on the shower pattern produced by electrons-positrons and protons in the HEIC. As described in section IV E, the HEIC has a good electron-to-proton (e-p) discrimination capability, which significantly reduces the proton background. However, discrimination between photons and electrons only depends on the ACD, which requires a minimum rejection fraction of $0.999$ for charged particles. Therefore, the ACD plays a critical role in identifying photons and distinguishing them from charged particles in the VLAST detector.

\subsection{Silicon Tracker and low Energy gamma-ray Detector }
The STED of VLAST is designed to fulfill three primary functions. Firstly, it reconstructs particle trajectories with an accuracy exceeding 120 $\mu \rm m$ for the majority of incident particles. Secondly, it determines the charge of cosmic rays. Lastly, it converts incoming photons into electron-positron pairs and detects both the photons and electrons resulting from Compton scattering. It consists of eight layers of Silicon Tracker and low Energy gamma-ray Detector modules. Each module consists of one CsI layer on the top and a Silicon Strip module at the bottom. The Silicon Strip module contains two layers of silicon strip detectors. The single-sided silicon strips are arranged in the x and y directions, with a pitch size of 120 $\mu \rm m$, defined as the distance between the centers of adjacent strips.
The top 6 layers of the 8-layer CsI have a thickness of 2~mm each, and the bottom 2 layers have a thickness of 4~mm each, resulting in a total radiation length of approximately $1~X_0$.
Each layer of CsI is assembled by joining numerous CsI (Na) scintillating crystal square panels, with each panel having sides measuring approximately 200 millimeters.
The upper and lower surfaces of each entire layer of CsI are tightly coupled to two layers of square wavelength-shifting (WLS) fibers arranged along the $x$ and $y$ axes, each having a cross-sectional side length of 2~mm. 
The WLS fibers absorb the blue scintillation light emitted from the CsI and transmit a fraction of the re-emitted red light to the SiPMs connected at their ends. 
The readout signals from the SiPMs can be used to infer the energy deposition and $x-y$ coordinates of particle interactions in the CsI panels.

 There are two main types of interactions for gamma-rays in the STED: Compton scattering and pair production. The dominant interaction depends on the energy of the gamma-rays. Compton scattering is predominant at energies below a few tens of MeV, while electron-positron pair production becomes dominant at higher energies. Different types of detectors are typically designed to capture gamma-rays in various energy bands. For instance, COMPTEL \cite{1993ApJS...86..657S}, Fermi-LAT \cite{2009ApJ...697.1071A}, and AGILE \cite{2009A&A...502..995T} focus on either the low or high energy range. In contrast, VLAST aims to detect gamma-rays in the energy band from 0.1~MeV to 1~TeV simultaneously, encompassing both interactions. Consequently, gamma-rays undergoing these two distinct interactions require different detection methods. The detection principle of VLAST is illustrated in  the lower-right panel of Fig.~\ref{fig:geomtry}.

For pair production events, the conceptual design of VLAST is similar to that of Fermi-LAT  and AGILE, but optimized for the lower energy band. The CsI serves a dual role as both a positron-electron pair converter and a detector for measuring the position and low energy deposition of interactions.
 This design ensures a high conversion efficiency of gamma-rays into electron-positron pairs. The CsI low-energy calorimeter does not reduce the multiple scattering effect of pair events. However, it can provide the deposited energy and position of fired tiles compared to tungsten plates, allowing for a more accurate assessment of energy and multiple scattering. Consequently, this design improves energy and angular resolution at low energies compared to Fermi-LAT and AGILE.

For the Compton scattering event, detection is challenging due to the scattered photons carrying away a significant portion of the momentum of the primary particle, especially when compared to pair production processes. VLAST must be capable of simultaneous measurement of two photons. Upon the entry of an incident photon into the detector, Compton scattering takes place within one layer of the tracker, giving rise to secondary photons and electrons. The photon scatters with an electron in the detector, transferring a fraction of its energy ($E_1$) to the electron. The scattered photon retains the remaining energy ($E_2$) and may undergo interaction with the CsI of the lower-energy detector behind it or enter the HEIC. The angle $\theta$ between the photon and the electron after Compton scattering can be calculated as
\begin{equation}
    \cos(\theta)=1- m_ec^2\left(\frac{1}{E_2}-\frac{1}{E_1+E_2}\right),
    \label{comp-equ}
\end{equation}
where $m_ec^2$ is the mass-equivalent energy of electron.

If only the angle of Compton scattering can be reconstructed, the incident direction of the photon can only be located on a circle on the sky, which is referred to as an ``event circle". These types of events are known as ``untracked" events. The width of the circle is related to the accuracy of the detector in measuring the direction of the scattered photon and the energy of the scattered electron. If the direction of the scattered electrons can also be measured, the ``event circle" becomes an ``event arc". These events are referred to as ``tracked" events, and the length of the arc reflects the accuracy of the measurement of the scattered electrons' direction. However, measuring such events is difficult for VLAST because the scattered electrons are easily absorbed by the CsI low-energy detector.

\begin{figure}[!htb]
\centering
\includegraphics[width=0.47\textwidth]{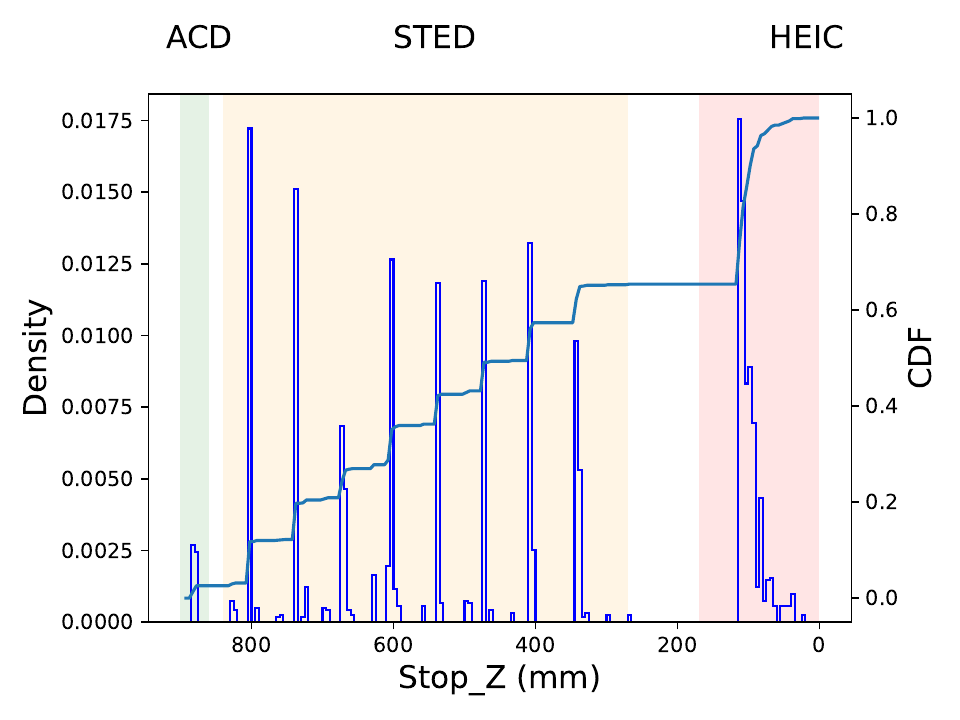}
\caption{The position distribution of 50 GeV photons converting into positron-electron pairs. From left to right, the peaks correspond to ACD, STED, and HEIC, respectively. The cumulative distribution  function (CDF) of conversion probabilities is displayed by the $y$-axis on the right side of the figure.}    
\label{fig:stopz}
\end{figure}

\begin{figure}[!htb]
\centering
\includegraphics[width=0.47\textwidth]{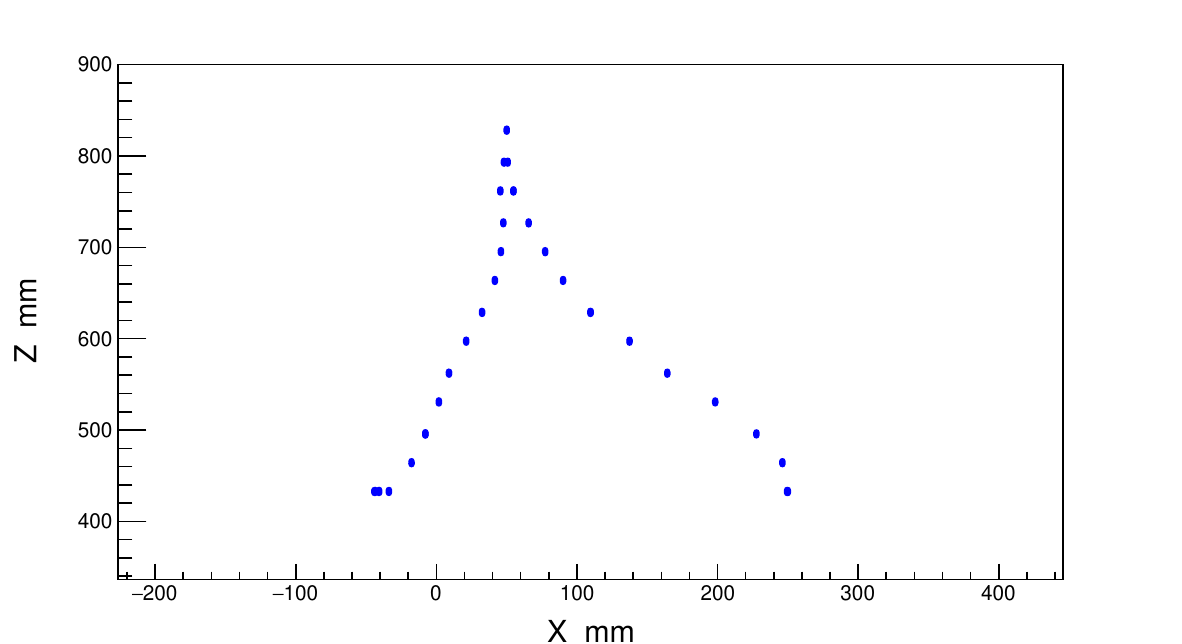}
\caption{Illustration of the hits in the STED for a 50~MeV gamma-ray photon ($x-z$ plane). The photon converts into an electron-positron pair in one of the CsI plates.}
\label{fig:tracker}
\end{figure}

The STED has a total thickness equivalent to one radiation length, which allows for a 65\% conversion rate of high-energy photons at normal incidence into electron-positron pair. Fig.~\ref{fig:stopz} illustrates the conversion position. Fig.~\ref{fig:tracker} provides an orthogonal cut view of a 50~MeV gamma-ray event detected in the tracker, demonstrating the generation of electron-positron pair in the CsI crystal and the gradual deviation of the electron and positron from their original path due to multiple scattering. Recoil particles cause the scattered signals. This inverted “V” signature is helpful in rejecting the much larger background of charged cosmic rays.

The main design challenge of STED is to balance the electron positron pair conversion efficiency with angular resolution in low-energy range. The angular resolution of low-energy photons is greatly affected by multiple scattering, which is dependent on $1/E$. It is necessary to design thinner CsI conversion material to reduce multiple scattering effects. However, this approach would reduce the conversion efficiency of low-energy photons. To resolve this trade-off, CsI is divided into two types:  thin layers are placed in the front section of the STED to ensure angular resolution for low-energy photons, while  thick layer are placed in the back to maintain high conversion efficiency for high-energy photons. For low energy photons of about 100~MeV, the direction is mainly determined by the two leading measurement points. In order to accurately reconstruct the direction of the incident photons, the first two measurement points after the photon transition point are crucial. To reduce the multiple scattering effects caused by the CsI plate and the support material, the efficiency of each measurement layer should be close to 100\%. So the spaced placement of CsI planes with  silicon strip planes allows precise measurement of the positive and negative electron traces. The thick CsI plane placed at bottom increases the efficiency of photon conversion at lower energies. It increases the statistics of low-energy photons. Studying the time-domain variation of photons is essential, despite the reduced angular resolution of photons. 


 The other challenge is balance the detector's field of view (FoV) and  angular resolution. For very high-energy photons, the impact of multiple scattering on angular resolution becomes less critical. The primary limitation on angular resolution stems from the ratio between the width of the silicon strip and the thickness of the silicon tracker detector. Ideally, finer silicon strips yield improved angular resolution. However, this comes with increased complexity in the fabrication process and higher electronic power consumption. On the contrary, increasing the thickness of the STED leads to a reduction in the detector's FoV. It also shifts the detector's center point upwards. STED was designed with this in mind.




\subsection{High Energy Imaging Calorimeter}
The HEIC has two main purposes: First, to measure the deposited energy of particles resulting from the interaction of incident photons. Second, to image the shower development profile which is used to reject cosmic ray  background  and also provides an estimate of the energy leakage fluctuations in the shower. To achieve high energy resolution, we have adopted the design of the BGO calorimeter on board the DAMPE \cite{2017APh....95....6C}. The HEIC is comprised of four identical modular towers arranged in a 2$\times$2 array, each tower has 416 BGO crystals divided into 4 layers with each adjacent layer placed orthogonally. Each BGO crystal is 2.5 cm $\times$ 2.5 cm $\times$ 1.4 m in size. The total vertical depth of HEIC is 18 radiation lengths or 1 nuclear action length, with oblique incidence events experiencing higher radiation lengths. The total effective detection area of HEIC is at last 2.8 m $\times$ 2.8 m. Signals from the BGO bars are read out at both ends by photomultipliers, the different dynodes of the photomultiplier tubes are responsible for different energy bands,  enabling VLAST to encompass a wide energy range from MeV to TeV. The left/right light asymmetry provides a measure of the energy deposit's position along the bar. Thus, each fired BGO crystal provides x, y, and z coordinates of the shower, and the shape of the shower can be inscribed in 3 dimensions. The inscribing pixels are directly related to the dimensions of the crystals. The shower axis also provides a rough track that can be used as a seed for the STED track reconstruction. Energy leakage is inevitable for grazing incidence and very high energy photons since the calorimeter has limited  radiation length. The leakage energy can be estimated by the shower profile from beam tests or simulations. Taking energy leakage correction into account significantly enhances the energy resolution of these events.

\section{Simulation and event reconstruction }\label{sec.III}
\subsection{Simulation}
\begin{figure}[!htb]
\centering
\includegraphics[width=0.47\textwidth]{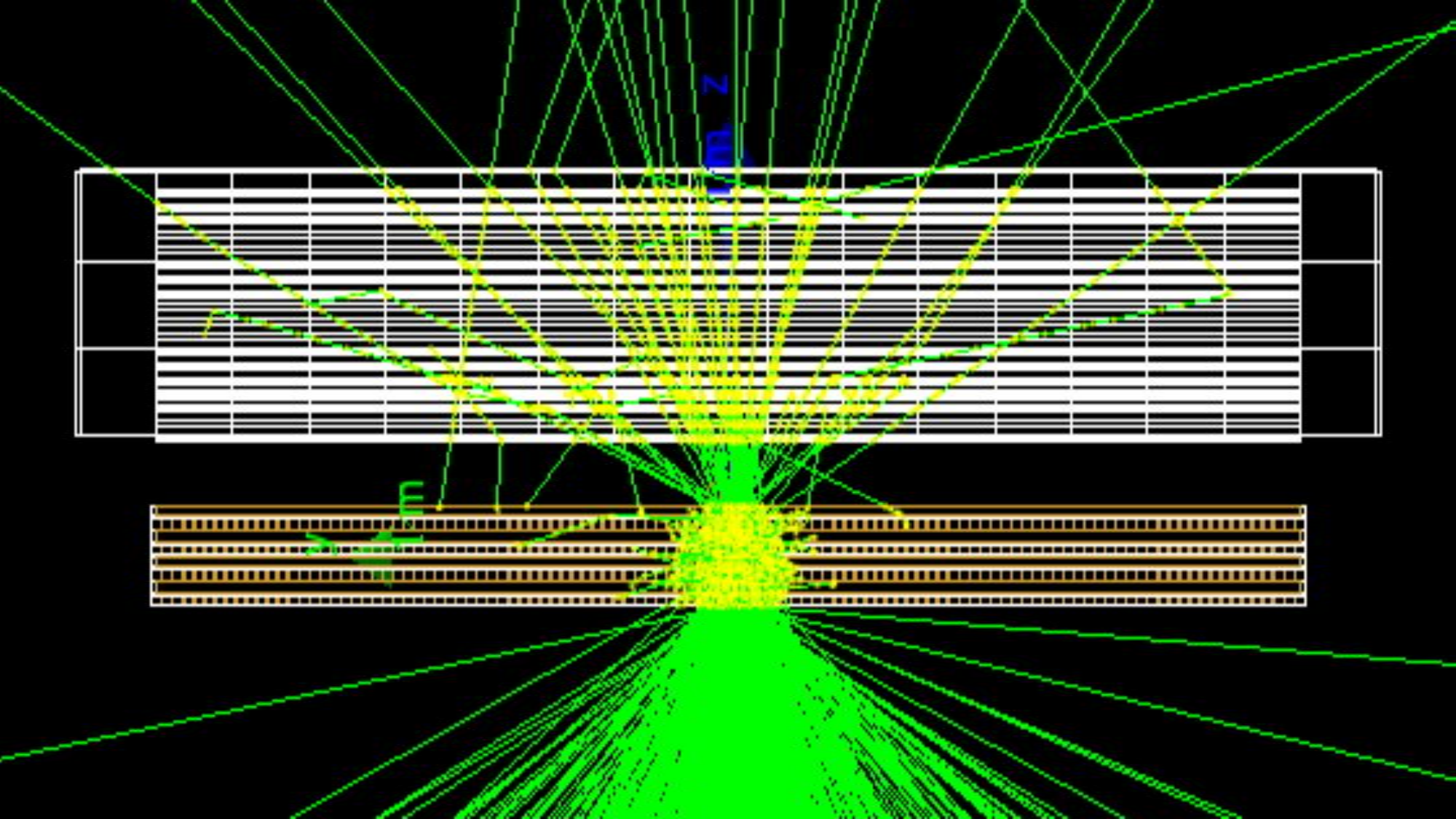}
\caption{Simulated shower particle tracks for a normal-incident 50~GeV photon. The backsplash particles would produce self-veto in the ACD.}
\label{fig:backsplash}
\end{figure}

To validate and optimise the VLAST design concepts, we have developed a simulation framework including incident particle definition, detector geometry definition, physical interaction list, digitization, and  event reconstruction analysis system. The framework is based on GEANT4, which is a publicly available toolkit developed in C++ for simulating the interactions between particles and matter. It includes numerous physical models of particle-matter interactions and is widely used in various fields such as high-energy physics, accelerator physics, space science, and medicine \cite{2016NIMPA.835..186A,2006ITNS...53..270A,2010EPJC...70..823A,2016AcASn..57....1H,2022icrc.confE..82J}. The gamma-ray source is set up as a sphere with an energy range of 0.1~MeV to 1~TeV. 
We use the GEANT4 FTFP\_BERT physics list to simulate incident particles with energies greater than 10 MeV. For 0.1--10 MeV gamma rays, we substitute the electromagnetic physics with G4EmStandardPhysics\_option4. This option introduces a more accurate model of Compton scattering and low-energy electromagnetic interactions, albeit at the cost of significantly increased computation time. We used the geometry module of GEANT4 to build the whole detector system include ACD, STED and HEIC. The constructed VLAST geometry with a normal-incident 50~GeV gamma-ray is shown in Fig.~\ref{fig:backsplash}. The output simulated data of different sub-detectors are digitized with the typical electronic noise of their corresponding readout systems\cite{he2023advances}. 

\subsection{Trigger design}

\begin{table*}[!htb]
\centering
\begin{tabular}{|c|c|c|c|c|c|c|c|}  
\hline  
 &  CsI-hit & HEIC-hit  & HEIC-HE  &HEIC-LE  & HEIC-MIP & STED  &ACD \\ \hline  
 MeV-Gamma&     1& 0 & $ \times$& $\times$ & $\times $& $\times $& 0 \\ \hline  
 GeV-Gamma&   $\times$ & 1 & 0 & $\times$ & $\times$ & 1 & 0 \\ \hline 
 LE&     $\times$ & 1 & 0 & 1 & $\times$ & 1 & 1 \\ \hline  
 HE&     $\times$ & 1 & 1 & $\times$ & $\times$ & $\times$ & $\times$ \\ \hline  
 Calibration&  $\times$ &1  & 0 & $\times$ & 1 & 1 & 1 \\ \hline  
\end{tabular}
\caption{Trigger logics of VLAST, including MeV-Gamma, GeV-Gamma, LE, HE and Calibration, and their corresponding requests (1: required; 0: excluded; $\times$: either). }
\label{tab:trigger}
\end{table*}

The preliminary trigger logic for scientific data collection consists of five trigger engines, including MeV-Gamma, GeV-Gamma, low energy (LE), high energy (HE), and Calibration, as shown in Table~\ref{tab:trigger}. Each sub-detector provides one or more trigger requests as detailed in the following list:
\begin{itemize}  
\item ACD: energy $>0.8$ MeV, corresponding to 0.4 MIP (minimum ionization particle); 
\item STED: three consecutive layers are on fire;
\item CsI-hit: two layers are on fire;
\item HEIC-hit: $>5$~MeV for the first or second layer; 
\item HEIC-HE: $>5$~MeV, $>500$~MeV, $>500$~MeV, $>500$~MeV for the first 4 layers;
\item HEIC-LE: $>5$~MeV, $>5$~MeV, $>50$~MeV, $>50$~MeV for the first 4 layers; 
\item HEIC-MIP: $>5$~MeV for the 1st, 2nd, 7th, and 8th layers.
\end{itemize}
The MeV-Gamma, GeV-Gamma, and HE trigger logics are combined to reserve gamma-ray events from sub-MeV up to multi-TeV. The comprehensive trigger efficiency is 45\%, 90\% and $>$95\% at 1~MeV, 1~GeV and above 10~GeV, respectively. The LE trigger logic (with a large pre-scale factor) is designed to reserve GeV cosmic-rays for performance validation on orbit. The Calibration trigger logic is for unit calibration of HEIC.

\subsection{Event reconstruction}
We utilized different algorithms to reconstruct the trajectories and energies of pair production and Compton scattering events, respectively. The Kalman filter was used to reconstruct the trajectory of the pair production event after charge sharing. The Compton scattering formula was adopted for the reconstruction of the direction of photon incidence. The longitudinal development of the electromagnetic shower was used to reconstruct the energy.


Before trajectory reconstruction, the process of charge sharing between silicon strips should be considered. This is mainly due to diffusion during charge collection and capacitive coupling. The drifting of electron-hole pairs along magnetic field lines causes the size of the charge cloud to increase, while capacitive coupling occurs on the coupling strip between the two readout strips. The charge sharing parameters come from DAMPE, which were obtained through a beam test \cite{2017APh....95....6C,2019NIMPA.935...24Q,2015ChPhC..39k6202D,2016ChPhC..40k6101Z,2018NIMPA.886...48Q,2016NIMPA.831..378A}. The charge sharing algorithm is
\begin{equation}
    E_{k}=E_{i}+0.023\cdot(E_{i-2}+E_{i+2})+0.0021\cdot(E_{i-4}+E_{i+4}), 
\end{equation}
if the strip is a readout strip, and
\begin{equation}
    \begin{aligned}
    E_{k}=& E_{i}+0.305\cdot(E_{i-1}+E_{i+1})+0.062\cdot(E_{i-3}+E_{i+3})\\
          &+0.012\cdot(E_{i-5}+E_{i+5}) +0.0024\cdot(E_{i-7}+E_{i+7}).
    \end{aligned}
\end{equation}
if the strip is a float strip.
After the process of charge sharing, we combine the signals from neighboring silicon strips with a signal/noise ratio larger than four (19 keV) into a cluster and then take the energy-weighted center as the position of the cluster. These clusters are then used in the subsequent analysis of trajectory reconstruction.

In this study, the Kalman filtering algorithm was employed to reconstruct the trajectory of the pair production event, a technique widely used in particle physics experiments \cite{1987NIMPA.262..444F,2002NIMPA.486..639C,2003OcDyn..53..343E}. When a gamma-ray converts into a pair in the CsI plane, the direction of the resulting electron/positron will experience a shift due to the multiple scattering effects during propagation. Hence, understanding the impact of multiple scattering in the trajectory reconstruction process becomes crucial. The Kalman filtering algorithm excels in assessing and compensating for both multiple scattering errors and measurement errors. This algorithm comprises three primary processes: prediction, filtering and smoothing. The track direction in the $k$-th layer is used to predict the hit position on the $(k+1)$-th layer. Subsequently, this predicted hit position of the $(k+1)$-th layer is adjusted using the measured hit. The evolution of the state vector is
\begin{equation}
    X_{k}=F_{k-1}X_{k-1} + W_{k-1},
\end{equation}
where $X_{k}$ is the state vector incorporating position and momentum information in the $k$-th layer, $F_{k-1}$ is the propagation of the trail from the $(k-1)$-th layer to the $k$-th layer of trail detector, and $W_{k-1}$ is the random noise of system.  In dense media, tracking particles is subject to random noise from multiple scattering, energy loss, or other physical processes that alter their trajectory. The trajectory offset caused by multiple scattering can be expressed as: 
\begin{equation}
    \sigma=\frac{13.6}{\beta c p}\sqrt{L/L_{\rm r}}[1+0.038\ln(L/L_{\rm r})],
\end{equation}
where $L$ is the thickness of CsI tile, $L_{r}$ is the radiation length of CsI material, $\beta c$ and $p$ are the velocity and the momentum (MeV) of electron-positron pair.
The measurement state vector is
\begin{equation}
    m_{k}=H_{k}X_{k}+V_{k},
\end{equation}
where $m_{k}$ is the quantities measured by the $k$-th layer, $H_{k}$ is the measurement matrix, and $V_{k}$ is the measurement error. Once a track has undergone the filtering process, it undergoes smoothing. Trajectory parameters are further refined from bottom to top, in contrast to the filtering process. For more details, please refer to Refs.~\cite{hernando1998kalman,2017ChA&A..41..455L,2002NIMPA.486..639C}.
 
The primary process for reconstructing the trajectory of MeV gamma-rays involves identifying Compton scattering events and determining the sequence of scattering points. 
The CsI hit found from top to bottom, with an energy deposition greater than 100 keV and no readout signals (less than 19 keV) from the silicon strips in the adjacent upper layer, is considered as the Compton scattering point.
Using the same method, additional isolated scattering points can be identified, and their scattering sequence as well as the probability of photoelectric effects can be analyzed through the magnitude of energy deposition.
Connecting the first and second interaction points with the highest probability, we can trace the trajectory of the photon after Compton scattering.
According to the Compton scattering formula (\ref{comp-equ}), to reconstruct the angle of Compton scattering can not determine the specific direction of the primary photon, but only a ring in the sky can be fixed. Multiple rings from the same source can be overlaid to locate the source.

For the electromagnetic shower induced by a high energy photon, the overall deposited energy in the CsI and BGO occupies most of the primary energy in despite of a small fraction of energy loss. However, with the increase of the incident energy, the longitudinal energy leakage becomes negligible and the energy correction is thus necessary to estimate the primary energy. The longitudinal segmentation of the CsI and BGO allows a fit of the longitudinal shower profile, which provides a good way to correct the longitudinal energy leakage \cite{2017NIMPA.856...11Y}. The longitudinal shower profile can be well described by a gamma-distribution formula as:
\begin{equation}
\frac{d E(t)}{d t}=E_0 \cdot \frac{(\beta t)^{\alpha-1} \cdot \beta \cdot e^{-\beta t}}{\Gamma(\alpha)},
\end{equation}
where $t=x/X_0$ represents the radiation length, $\Gamma(\alpha)$ is the gamma function, $\alpha$ and $\beta$ are the shape factor and scale factor, respectively. The depth of the shower maximum depends on $\alpha$ and $\beta$, as $t_{\rm max} = (\alpha-1)/\beta$. The value of $t_{\text{max}}$ is closely correlated with the energy leakage ratio, offering an effective approach for energy correction~\cite{2017NIMPA.856...11Y}.


\section{Expected performance of the VLAST}\label{sec.IV}

\subsection{Effective area}

The effective area reflects the detection efficiency of an instrument. It is equivalent to the area of an ideal absorber that detects the same number of events as the real detector, considering event selection and reconstruction, within the same time. The effective area can be calculated as \cite{2014arXiv1407.7631B}

\begin{equation}
A_{\mathrm{eff}}(E, \theta, \phi)=A_{\mathrm{geo}}(\theta, \phi) \varepsilon_{\mathrm{det}}(E, \theta, \phi) \varepsilon_{\mathrm{sel}}(E, \theta, \phi),
\end{equation}
where $A_{\mathrm{geo}}(\theta, \phi)$ represents the geometric cross-sectional area of the instrument in a specific direction characterized by angles $\theta$ and $\phi$,  $\varepsilon_{\mathrm{det}}(E, \theta, \phi)$ is the detection efficiency, and $\varepsilon_{\mathrm{sel}}(E, \theta, \phi)$ is the selection efficiency.
The detailed selection conditions for pair events include: 1) the trigger condition for particular event types is satisfied (MeV-Gamma, GeV-Gamma or HEIC-HE), 2) the trajectory in the STED can be reconstructed correctly, and 3) the deposited energy in the ACD block in the direction of the reconstructed trajectory is less than 0.5 MeV. The selection conditions for Compton events include: 1) no ACD readout exceeding 100 keV, 2) there are at least two CsI hits where the energy deposition at both points is greater than 100 keV and no readout signal (less than 19 keV) in the adjacent upper and lower layers of silicon tracker detector, and 3) the energy deposition in the bottom CsI layer is smaller than 50 keV (to ensure that the majority of photon energy is deposited in the STED).

\begin{figure*}[!htb]
\centering
\includegraphics[width=0.45\textwidth]{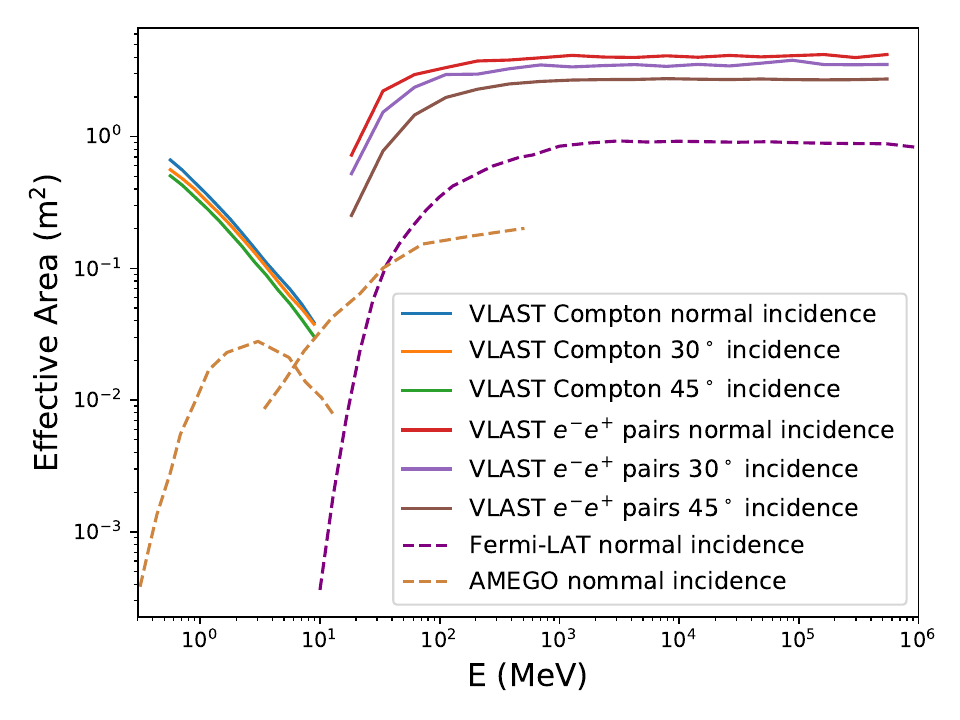}
\includegraphics[width=0.45\textwidth]{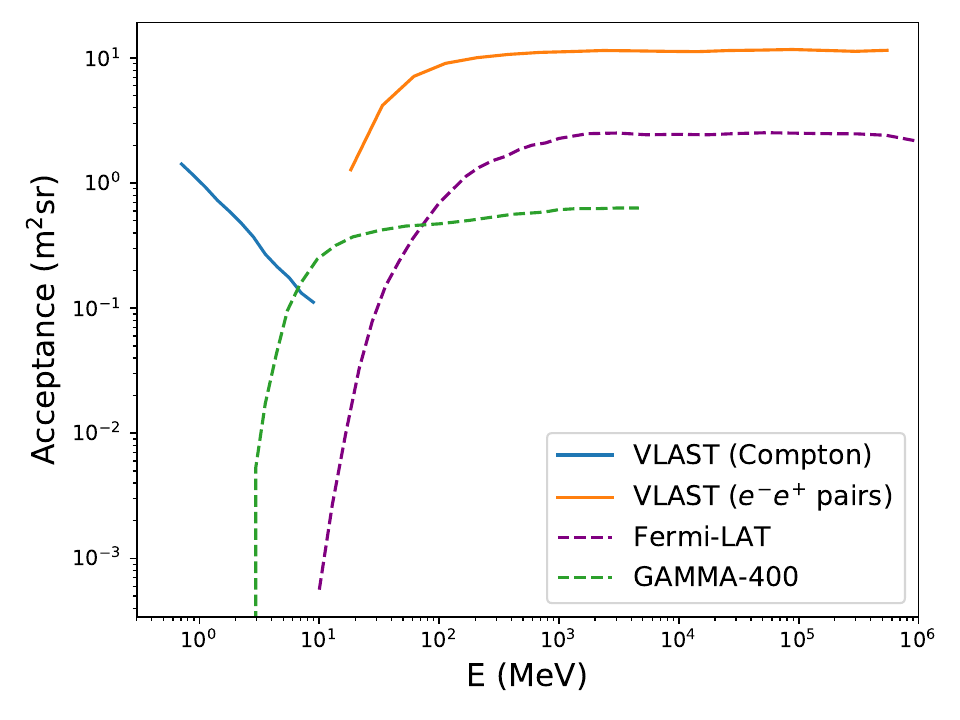}
\caption{Left: effective areas of VLAST for photon events with normal incidence, $30^\circ$, and $45^\circ$ incident angles, compared with those of AMEGO \cite{2017ICRC...35..783C,2019BAAS...51g.245M} and Fermi-LAT \cite{Fermi_webpage}. 
Right: acceptance of the VLAST for gamma-rays of different energies, compared with results of GAMMA-400 \cite{2022AdSpR..70.2773T} and Fermi-LAT \cite{Fermi_webpage}.}
\label{fig:my_acc_eff}
\end{figure*}

\begin{figure}[!htb]
\centering
\includegraphics[width=0.47\textwidth]{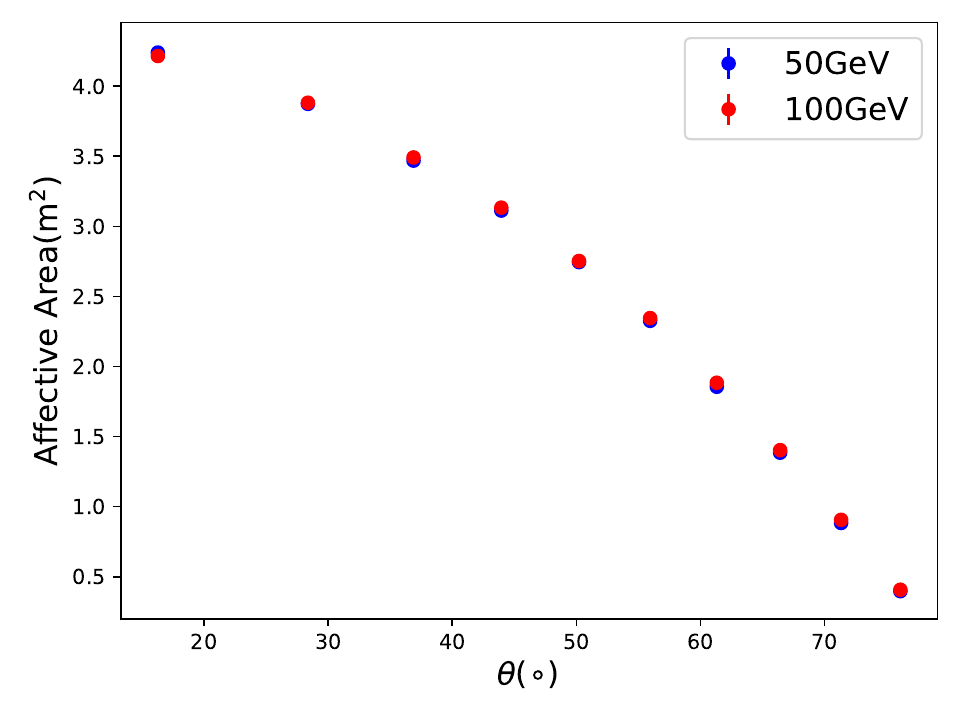}
\caption{Effective area for gamma rays with different incident angles.}
\label{fig:AEFF_Theta}
\end{figure}

The effective area of VLAST is shown in Fig.~\ref{fig:my_acc_eff}. For Compton scattering events, the effective area is several thousand $\rm cm^2$, surpassing the COMPTEL detectors' effective area (10-50 $\rm cm^2$) by one or two orders of magnitude. It also has a larger effective area comparable with the planned MeV detectors, such as e-ASTROGAM, AMEGO, GECCO, and XGIS-THESEUS detectors. Furthermore, the effective area for pair production events exceeds $\rm 4~m^2$ above 1~GeV with normal incidence, four times larger than that of the Fermi-LAT detector. However, the effective area slightly decreases for $30^{\circ}$  and $45^{\circ}$ incidence angle for both Compton scattering and pair production events. Fig.~\ref{fig:AEFF_Theta} depicts the effective area as a function of the particle incidence angle for different energies (50~GeV, 100~GeV). The effective area diminishes gradually with an increasing angle of incidence, and the maximum off-axis incidence angle can reach $70^{\circ}$. The FoV of VLAST is 2.5 sr.

The  acceptance of a gamma-ray detector refers to the portion of incoming gamma-rays that the detector is capable of capturing or detecting. A larger acceptance indicates that the detector can detect gamma-rays from a wider portion of the sky, thereby increasing its sensitivity to gamma-ray detection. The primary design concept of VLAST is to substantially enhance the detector's acceptance in order to capture more photons and achieve more precise measurements with increased statistical significance. So the  acceptance of VLAST must be sufficiently large. The acceptance is defined as the integral of the effective area over the solid angle, $G(E)=\int_{\Omega} A_{\mathrm{eff}}(E, \theta, \phi) d \Omega$. The  acceptance of VLAST is presented in Fig.~\ref{fig:my_acc_eff}. For Compton scattering events, the maximum value of the  acceptance is several $\rm m^2 sr$. As the energy increases from the MeV to GeV range, the acceptance gradually rises from 1.6 $\rm m^2 sr$ to 12 $\rm m^2 sr$. In the design of VLAST, CsI crystal plates replace the tungsten foils used in many other gamma-ray detectors to enhance  pair conversion. This substitution significantly improves the acceptance in the MeV energy range compared to the Fermi-LAT. Given its substantial size, the  acceptance of VLAST reaches 12 $\rm m^2 sr$ in the energy range above GeV.

\begin{figure}[!htb]
\centering
\includegraphics[width=0.47\textwidth]{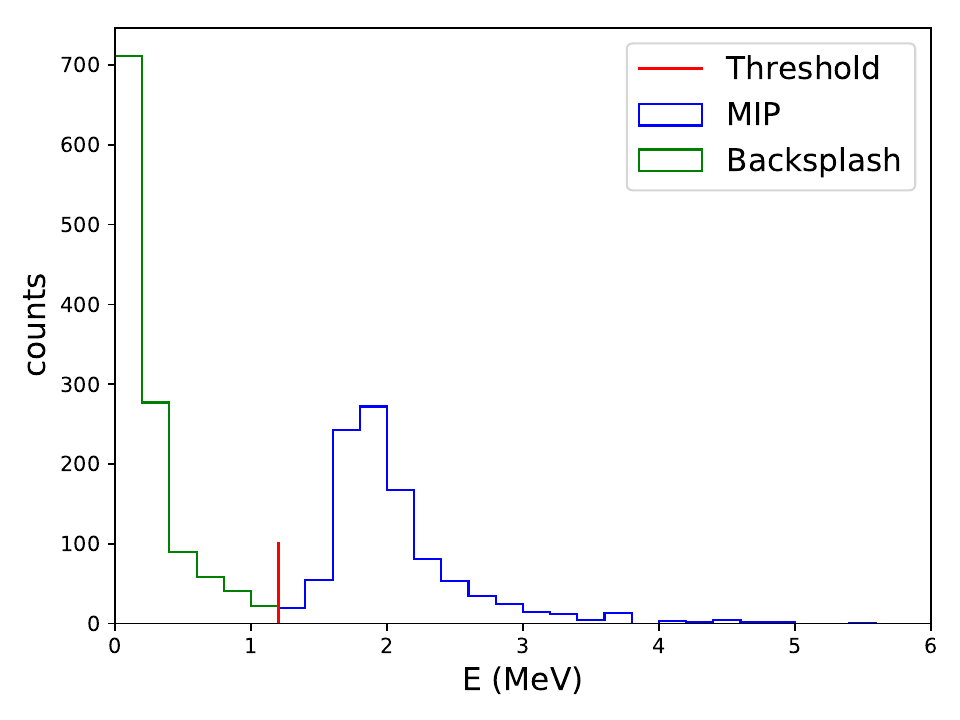}
\caption{Energy deposition distribution of MIPs and backsplash effect in ACD. The optimized threshold that balances the high detection efficiency of charged primary particles and low contamination of backsplash is marked out in the plot.}
\label{fig:acd_con}
\end{figure}

\begin{figure*}[!htb]
\centering
\includegraphics[width=0.33\textwidth]{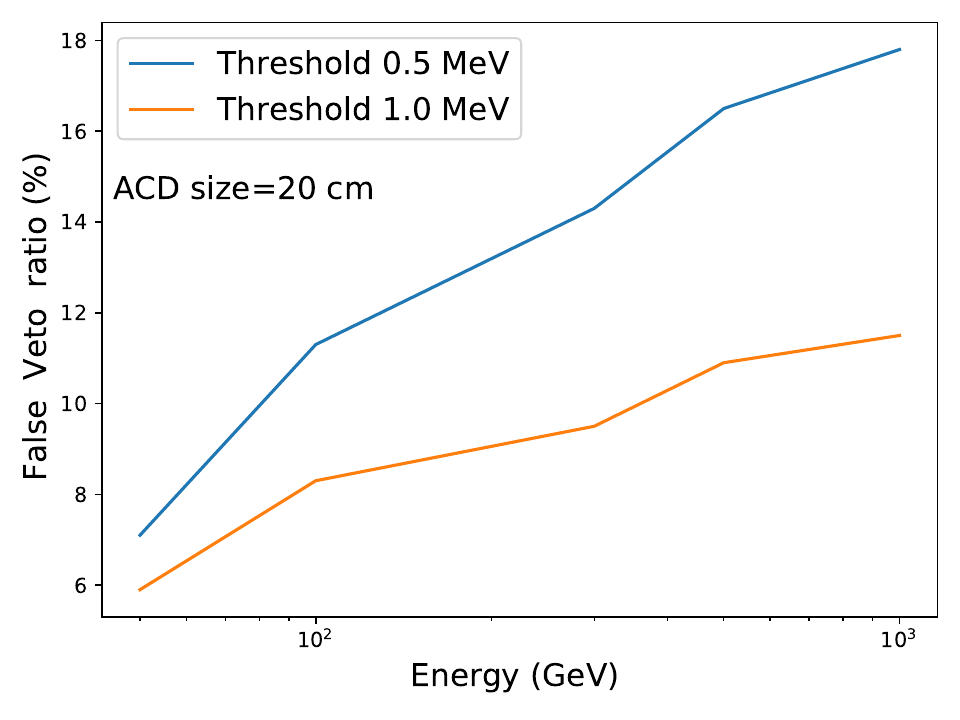}
\includegraphics[width=0.33\textwidth]{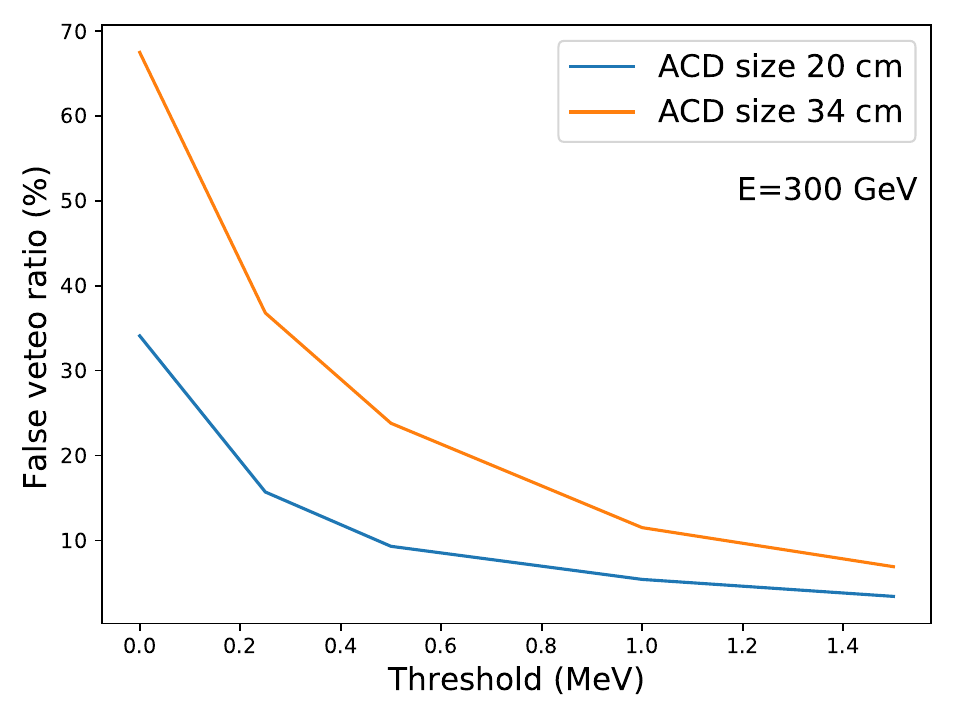}
\includegraphics[width=0.33\textwidth]{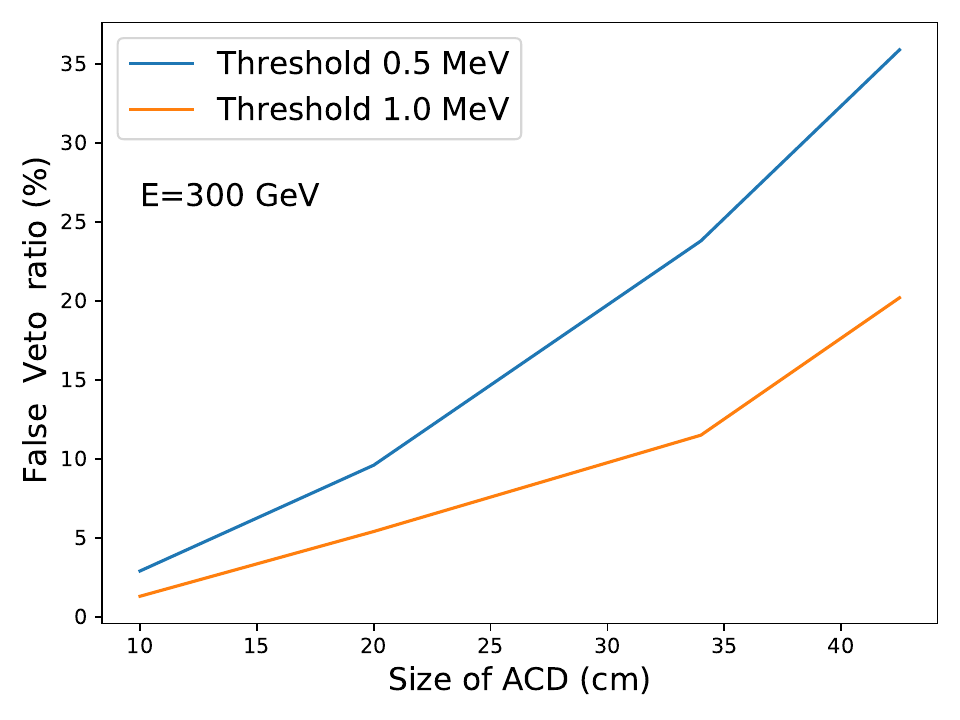}
\caption{False veto ratios as functions of incident energy (left), the ACD threshold (middle), and the ACD size (right). }
\label{fig:veto_threshold_energy_size}
\end{figure*}

\subsection{Self-veto}
The detailed settings of the ACD need to be optimized, including the detection threshold and the size of the ACD. The detection threshold of the ACD must strike a balance between the detection efficiency of charged particles and the suppression of backsplash effects. To improve the efficiency of detecting charged particles, which exhibit a Landau distribution in energy deposition within the ACD, the detection threshold should be set lower. However, in order to reduce weak signals from recoil particles, the detection threshold should be set higher. The conflict between these two requirements is illustrated in Fig.~\ref{fig:acd_con}.    
We utilized the false veto ratio of gamma-rays to determine the threshold. Gamma-ray photons might be misidentified as charged particles and subsequently rejected due to the backsplash effect. Even if photons do not leave a signal when passing through the ACD, the low-energy recoil photon could trigger the ACD. As a result, the tile in the direction of particle  could exhibit a signal, leading to the event being incorrectly identified as a charged particle. The false veto ratio is defined as $f=N_{\rm wrong}/N_{\rm total}$, where $N_{\rm wrong}$ are the events of photon being misidentified as a charged particle and $N_{\rm total}$ is the total number of incident photons. The false veto ratio as a function of energy is shown in the left panel of Fig.~\ref{fig:veto_threshold_energy_size}. The higher the energy of the photon, the greater the probability that it will be misclassified as a charged particle.  The middle panel of  Fig.~\ref{fig:veto_threshold_energy_size} presents the gamma-ray false veto ratio at different detection thresholds.  For the ACD size of $20\times 20 ~ \rm cm^2$, we find the optimized threshold value is 0.5~MeV where the gamma false veto rate is less than 15\%. The size of the ACD should be  also set reasonably.  Under ideal conditions, a smaller ACD block would result in better suppression of the backsplash effect. However, this also means more readouts and a higher power consumption. Thus, to achieve the desired performance under limited conditions, we conducted tests to evaluate the suppression of the backsplash effect for different ACD sizes. The results of these tests are presented in the right  panel of Fig.~\ref{fig:veto_threshold_energy_size}. An ACD tile size of $20\times20 ~ \rm cm^2$ is a reasonable choice when the gamma false veto rate is less than 15\%. The optimization of the ACD thickness is also necessary. The ACD needs to be of sufficient thickness to generate adequate fluorescence, ensuring effective detection of charged particles. However, it should not be excessively thick, as this could lead to a significant number of photons interacting within the ACD being misidentified as charged particles. In this context, we set the thickness of the ACD to 1 cm, the same as that used for Fermi-LAT. With this thickness, no more than 3\% of photons interact within the ACD.
The interactions of photons within the ACD are illustrated in Fig.~\ref{fig:stopz}. Moreover, it is advisable to keep the wrapping materials around the ACD, such as micrometeoroid shields and thermal blankets, thin. This precaution is important because charged cosmic rays, upon interacting with such materials, can generate secondary photons that contribute to a local photon background. This background can interfere with the accurate detection of gamma-rays, reducing detection efficiency.

\subsection{Angular resolution}

The angular resolution is another critical parameter of VLAST. A better angular resolution not only provides more accurate source positioning but also results in a sharper profile of the point source. This, in turn, leads to a smaller contribution from the background, thereby increasing the sensitivity to detect fainter sources. The angular resolution is defined differently for different interaction processes, namely Compton scattering or pair production processes. For Compton events,  the angular resolution is defined as the sigma value obtained from the Gaussian fit to the distribution of the minimum angular distance between the nominal source position and the reconstructed event circle. In the case of pair production events, it is defined as the radius of the circle that includes 68\% of the point spread function (PSF). The PSF is determined by the angular distance between the source and the reconstructed direction.

\begin{figure}[!htb]
\centering
\includegraphics[width=0.47\textwidth]{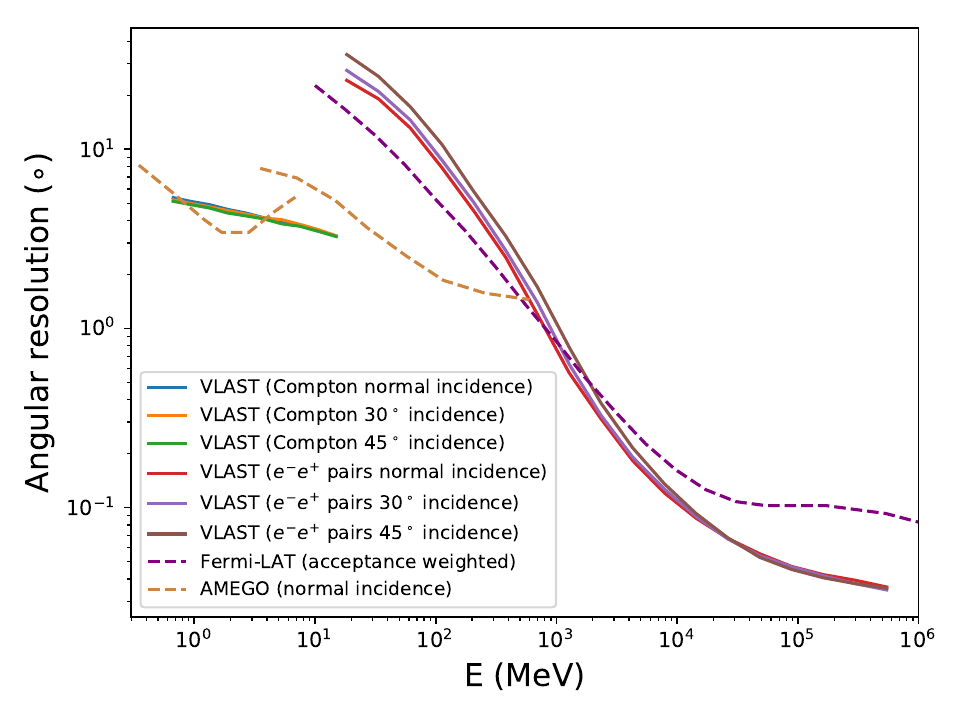}
\caption{Angular resolution (68\% containment) of VLAST for photon events with normal incidence, $30^\circ$ and $45^\circ$ off-axis angles, compared with results of AMEGO \cite{2017ICRC...35..783C,2019BAAS...51g.245M} and Fermi-LAT \cite{Fermi_webpage}.}
\label{fig:psf}
\end{figure}

The angular resolution of VLAST is depicted in Fig.~\ref{fig:psf}. To get these results, we use the same selection conditions as in Sec. IV A for the calculation of the effective area.  For Compton events, the angular resolution falls within the range of 4$^\circ$ to 8$^\circ$ which is comparable to AMEGO. Here, we present the angular resolution for untracked events only. As for pair production events, the angular resolution improves with increasing photon energy, owing to the reduced impact of multiple scattering effects at higher energies. Notably, at 10~GeV, the angular resolution reaches 0.2$^\circ$. However, between the energy range of 10 to $10^3$ MeV, the angular resolution is slight worse than that of Fermi-LAT.  In contrast, in the energy range above GeV, the angular resolution surpasses that of Fermi-LAT. This improvement can be attributed to our use of analog readout, whereas Fermi-LAT employs digital readout with silicon strips of roughly the same width.

\begin{figure*}[!htb]
\centering
\includegraphics[width=0.45\textwidth]{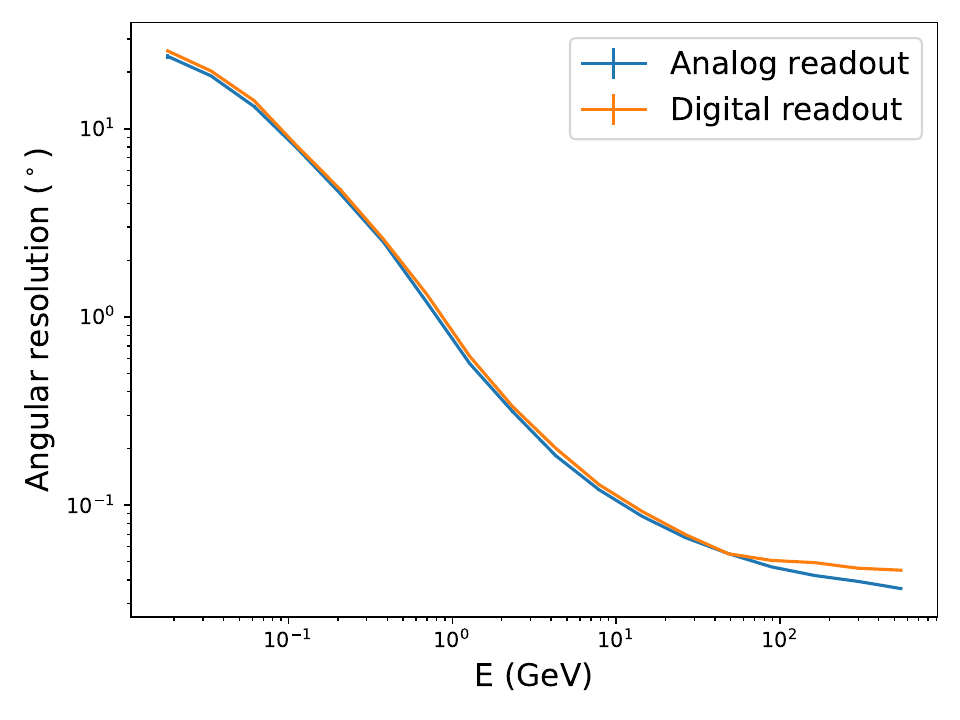}
\includegraphics[width=0.45\textwidth]{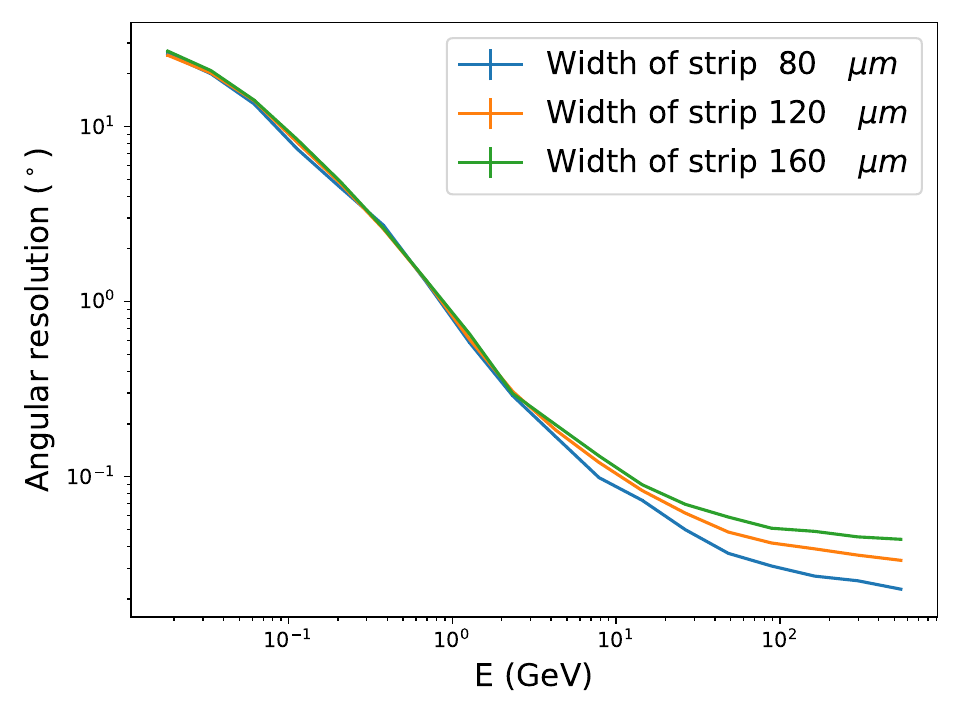}    
\caption{Angular resolution of VLAST for normal incident photons. The left panel shows the comparison between digital and analog readouts for 120 $\mu$m strip width, and the right panel compares different widths of silicon strips for analog readout.}
\label{fig:psf_digital_strip}
\end{figure*}

\begin{figure*}[!htb]
\centering
\includegraphics[width=0.45\textwidth]{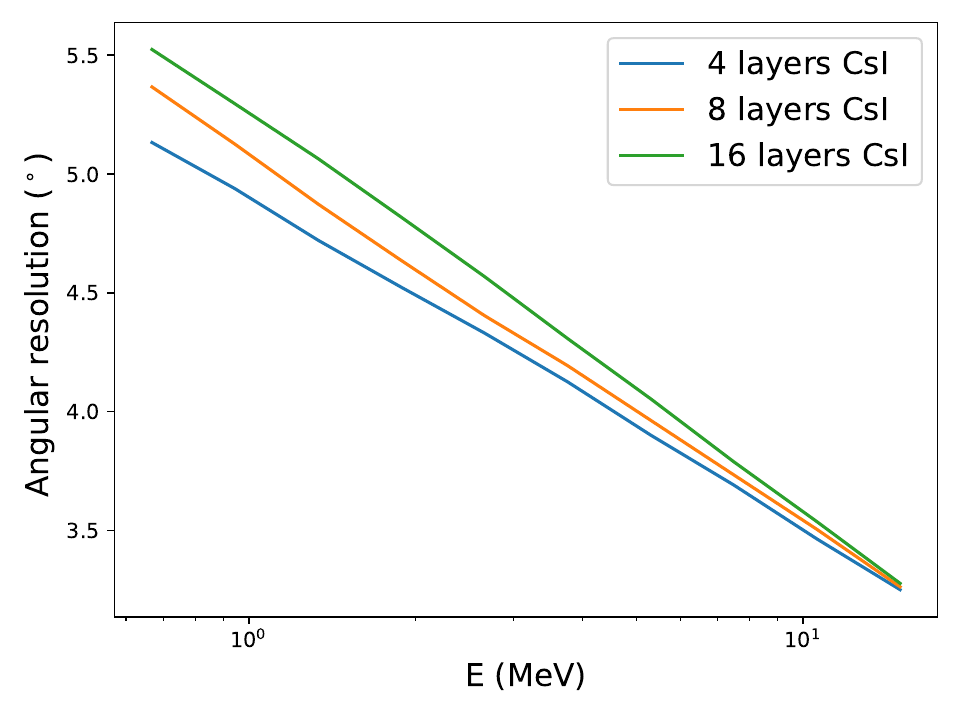}
\includegraphics[width=0.45\textwidth]{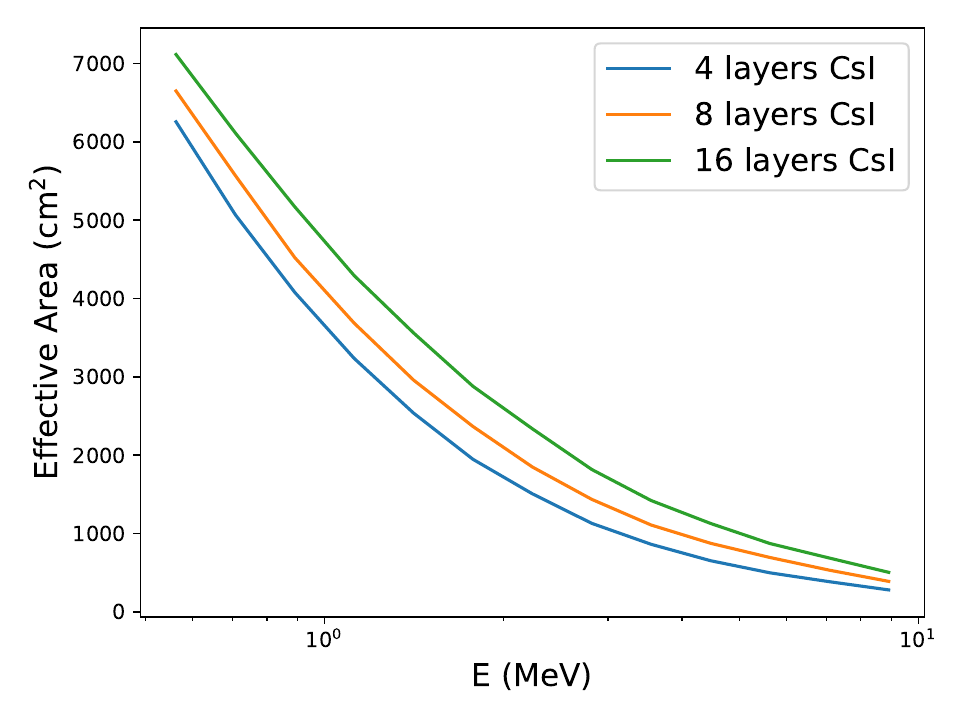}
\caption{Angular resolution (left) and effective area (right) of VLAST for different configurations of CsI layers for Compton events. Normal incidence and analog readout are adopted in the simulation.}
\label{fig:area_psf_layer_MeV}
\end{figure*}

As mentioned earlier, we have implemented a spaced readout scheme for the silicon strip readout to reduce power consumption. Despite this scheme, the VLAST still requires  a large amount of  electronic readouts, resulting in substantial power consumption. To tackle this issue, we have explored the use of digital readout, which records only the hit strips that exceed a signal threshold of 3$\sigma$ above the baseline noise as ``1", while unhit strips are recorded as ``0". This simplifies the electronics significantly and reduces power consumption. We define a cluster as a region of hit silicon strips. For digital readout, during cluster position reconstruction, the cluster's position is set to the geometric center, while for analog readout, the energy-weighted average position is used for cluster reconstruction. Subsequently, track reconstruction is performed using Kalman filtering. A comparison between analog and digital readout is presented in the left panel of Fig.~\ref{fig:psf_digital_strip}. In the lower energy range, the dominant error arises from multiple scattering, resulting in comparable angular resolution for both analog and digital readouts. However, in the higher energy range, measurement errors take precedence, and the geometric center of the cluster exhibits a significantly larger error compared to the energy-weighted center. This difference increases with energy, showing that analog readouts provide better angular resolution than digital readouts.

We investigate the impact of different widths of silicon strips on the PSF of VLAST for pair production events. Thinner strips offer higher position measurement accuracy and better angular resolution. However, they also necessitate more electronic readouts and result in higher power consumption. Therefore, a balance between angular resolution and power consumption is required. We conducted tests using three different widths: 80~$\mu \rm m$, 120~$\mu \rm m$, and 160 $\mu \rm m$. The results are shown in the right panel of Fig.~\ref{fig:psf_digital_strip}. Notably, there is no big difference in angular resolution at the lower energy band among the different widths. This is because the uncertainty of angular resolution is primarily attributed to multiple scattering effects within this energy range. However, as multiple scattering effects decrease at higher energy levels, measurement uncertainty become more prominent, and finer silicon strips lead to better resolution. As a benchmark, we chose a silicon strip width of 120~$\mu \rm m$, which gives an angular resolution of 0.05$^\circ$ at 50~GeV.

\begin{figure}[!htb]
\centering
\includegraphics[width=0.47\textwidth]{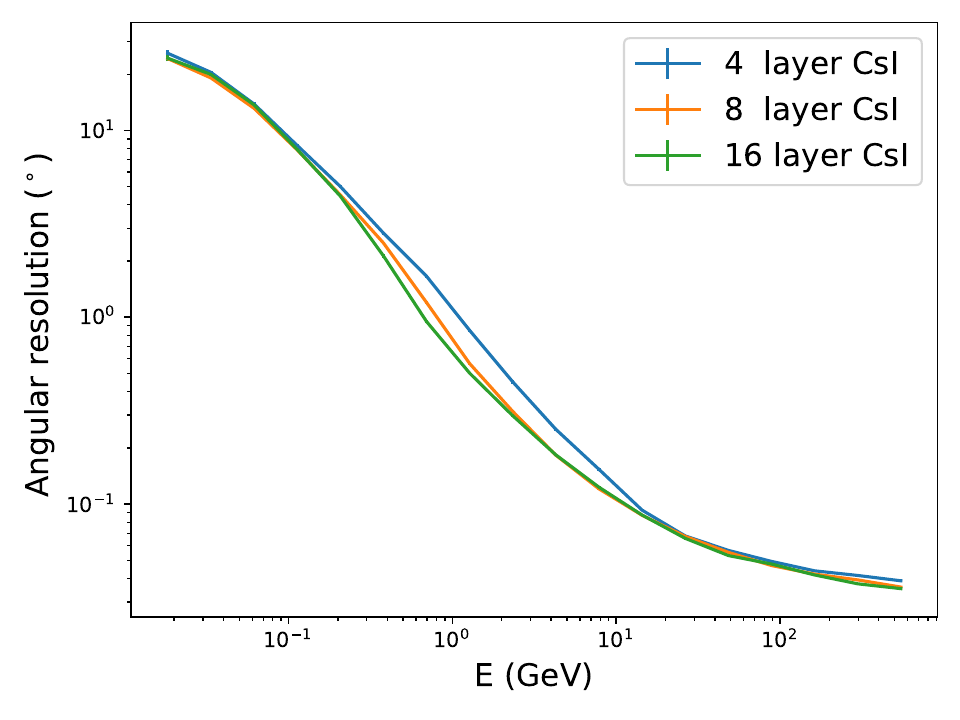}
\caption{Angular resolution of VLAST for different configurations of CsI layers for pair events.  Normal incidence and analog readout are adopted in the simulation.}
\label{fig:psf_layer_GeV}
\end{figure}

In this study, we investigated the effect of different CsI layers on angular resolution. In order to simplify the mechanical structure and ease the manufacturing and assembly process, the STED prefers thicker CsI plates with fewer layers. However, thicker CsI plates degrade the angular resolution because of the larger deflection angle due to multiple scattering effects. Therefore, the thickness of the CsI plate needs to be set appropriately.  We tested three different configurations with 4, 8, or 16 layers of CsI plates interleaved in the STED, while maintaining the same total radiation length to ensure the same efficiency of pair production. The effect of different layer settings on Compton scattered events  is shown in the left panel of  Fig.~\ref{fig:area_psf_layer_MeV}. Different layer settings have opposite effects on angular resolution and effective area. The angular resolution decreases with  number of CsI layers. Because 4 layers have thickest CsI that measured energy is most accurate, the corresponding reconstruction direction is the most accurate according to the Compton scattering equation. But, the effective area increases with  number of CsI layers. The 4-layer CsI configuration has the smallest effective area. For pair production events, the effect of different CsI layer configurations on angular resolution is shown in  Fig.~\ref{fig:psf_layer_GeV}. The angular resolution is not significantly different in the low energy band. The 16-layer configuration offers the best angular resolution at several GeV, due to the reduced multiple scattering in thinner CsI. However, at several hundred GeV, the difference becomes negligible as multiple scattering effects diminish, while measurement errors become more significant when using the same pitch of silicon strip. Different layer configurations have the same effective area because the total radiation length is the same. Based on comprehensive considerations, we determined that setting the number of CsI layers to 8 strikes the most suitable balance.

\subsection{Energy resolution}
\begin{figure}[!htb]
\centering
\includegraphics[width=0.47\textwidth]{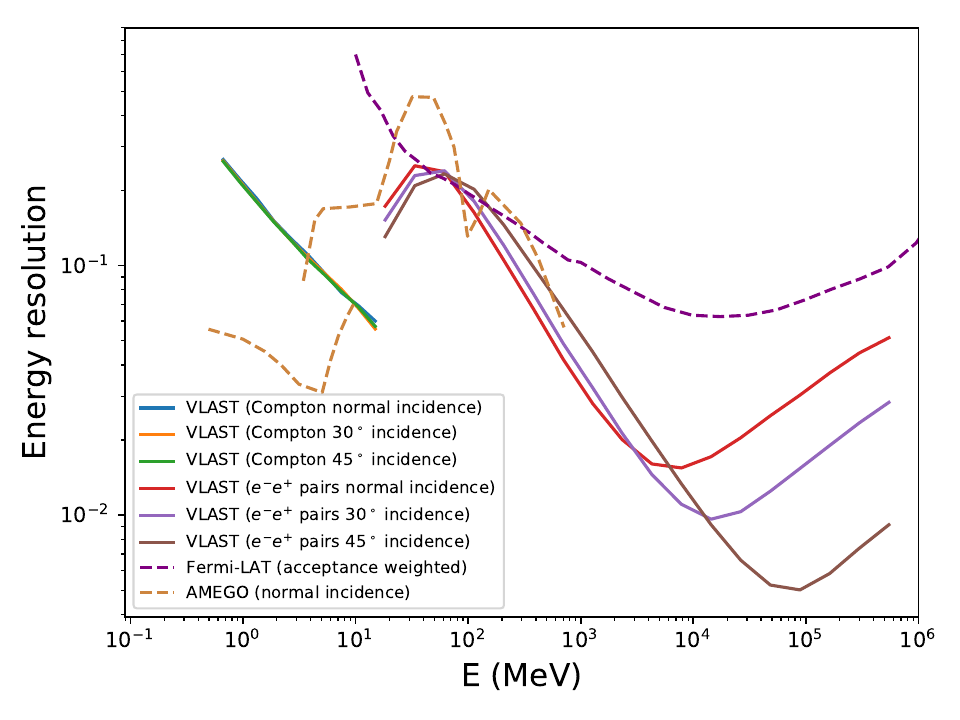}
\caption{Energy resolution of VLAST for Compton and pair events with normal incidence, $30^\circ$ and $45^\circ$ incident angles, compare with results of AMEGO \cite{2017ICRC...35..783C,2019BAAS...51g.245M} and Fermi-LAT \cite{Fermi_webpage}.}
\label{fig:E_res}
\end{figure}

The energy resolution is another critical performance of the VLAST. One kind of the scientific objectives of the most interests for the VLAST are line structure in the spectra produced from astrophysical phenomena, such as de-excitation nuclear gamma-ray line emission from low energy cosmic ray and  the gamma-ray  line from  possible annihilation of dark matter particles \cite{2013PhRvD..88h2002A,2022SciBu..67..679L,2013JCAP...04..017C}. The ability of detecting the line structure in the gamma-ray spectrum relies extremely on the energy resolution. Here we analyze the energy resolution of the STED and HEIC  with the above mentioned configuration. For pair events, first, the trigger conditions are MeV-Gamma, GeV-Gamma and HEIC-HE. Second, In STED, kalman filtering can reconstruct the photon trajectory. Third, Most of the electromagnetic shower is contained in HIEC. The energy deposition obtained from simulation, considering statistical fluctuation error and electronic noise (based on DAMPE \cite{2015NIMPA.780...21Z, 2015ITNS...62.3117F,2022NIMPA102966453Z,2019NIMPA.922..177W}), is employed for the energy resolution analysis. For high-energy events with more energy leakage, the energy resolution is obtained after energy correction. For the Compton events, the energy is measured exclusively by the CsI calorimeter, and the uncertainty of measurements is primarily considered in four aspects: quantum fluctuations of CsI scintillation light yield, transmission efficiency of the WLS fibers, photon detection efficiency and electronic noise of SiPMs, and accuracy of energy calibration. The process of energy measurement is simulated by randomly sampling each energy deposition of CsI, utilizing specific parameters initially estimated from previous experiments. The energy resolution is obtained by Gaussian fitting the proportion of the sample results to the simulated real energy.
The energy resolution is shown in Fig.~\ref{fig:E_res}. 
For pair production event, There is a peak around 100 MeV, because the deposition body shifts from low-energy calorimeter to  HEIC.  As the energy particles entering the calorimeter increases, the energy resolution, which becomes better, is able to reach $2\%$ at a few tens of GeV. Subsequently, become weak due to energy leakage. The energy resolution of the $30^{\circ}$ and $45^{\circ}$ incidence event are better than normal incidence event because the radiation length of path are longer.  In comparison to Fermi-LAT, which has an approximate radiation  length of 8.6, VLAST exhibits better energy resolution across the entire energy band due to its deeper calorimeter except around 100 MeV.

\begin{figure*}[!htb]
\centering
\includegraphics[width=0.45\textwidth]{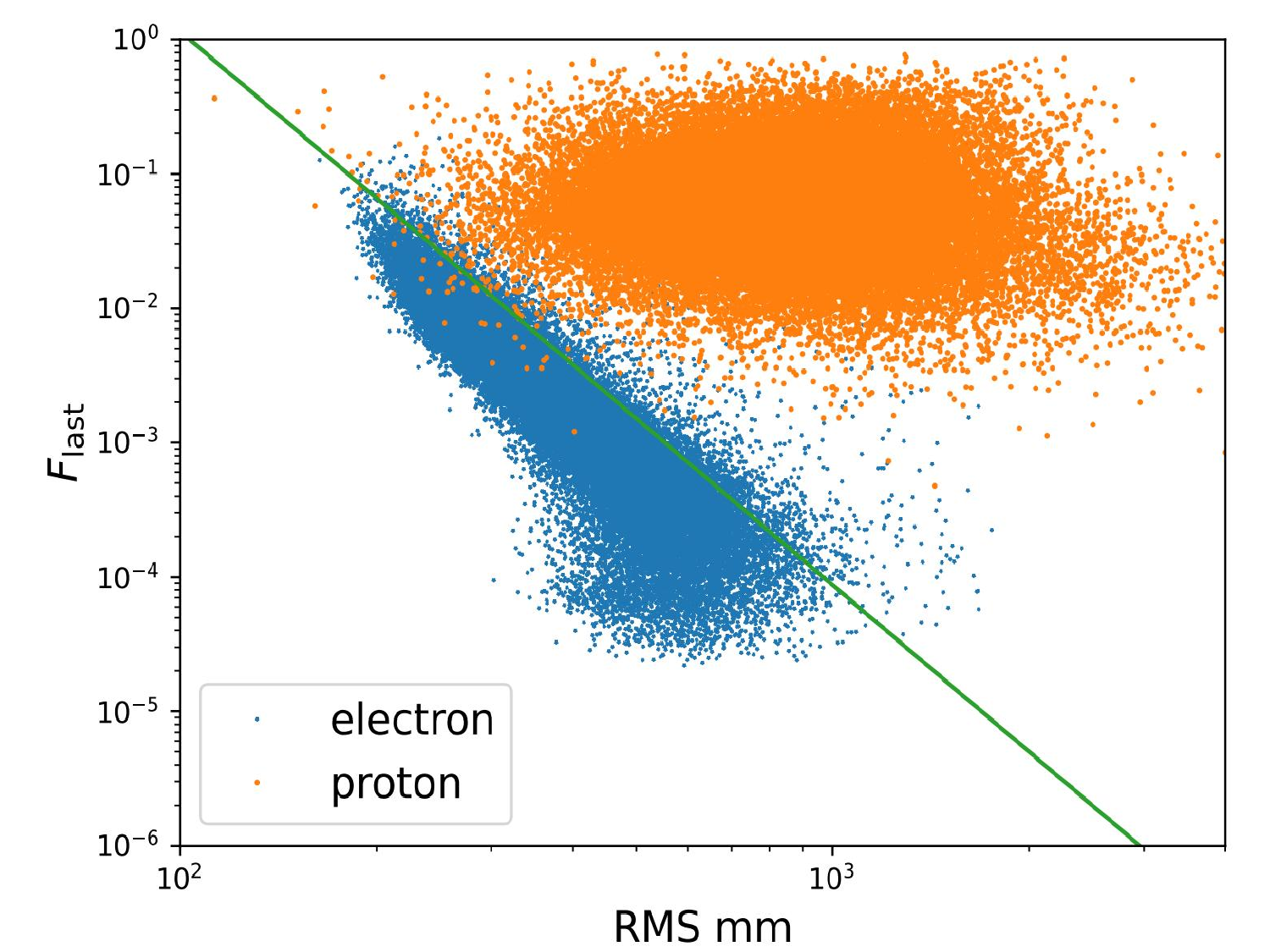} 
\includegraphics[width=0.45\textwidth]{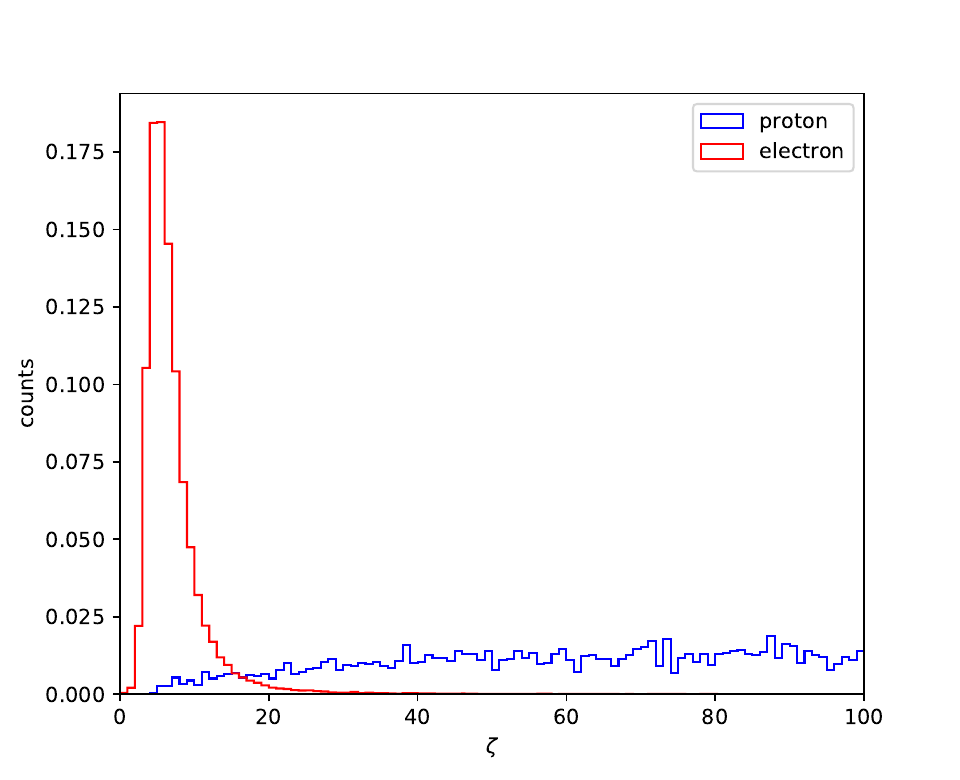}
\caption{Left: Distributions of the lateral and longitudinal development of showers for electrons (blue dots) and protons (orange dots) in the HEIC. The green line illustrates the selection condition to seperate these two populations. Right: One-dimensional distributions of $\zeta$ variables of electrons and protons.}
\label{fig:e-p}
\end{figure*}

\subsection{Electron and proton discrimination}

VLAST requires a strong capability to reject charged particle background in order to detect photons more effectively. As previously mentioned, the proton flux is five orders of magnitude higher than that of photons, making it difficult to detect gamma rays in the presence of such a large background of charged particles. In addition to the ACD, the HEIC plays another crucial role in background rejection by distinguishing protons from photons based on the difference between hadronic and electromagnetic showers. To quantify the e-p discrimination ability of the VLAST, we use the the $\zeta$ parameter as used for DAMPE \cite{2017Natur.552...63D}, which is defined as
\begin{equation}
\zeta=F_{\rm last} \times (\Sigma_{i}{\rm RMS}_{i}/\rm mm)^4/(8\times10^6),
\end{equation}
where $F_{\rm last}$ is the ratio of energy deposition in the last layer to the total energy deposition in the HEIC. The energy deposition in the last layer reflects the differences between hadronic and electromagnetic showers most significantly because BGO crystals have a large nuclear interaction length to radiation length ratio. The quantity ${\rm RMS}_{i}$ is the root-mean-square value of the energy deposited hits' position in the $i$-th layer:
\begin{equation}
{\rm RMS}_{i}=\sqrt{\frac{\Sigma_{j}( x_{j,i}-x_{c,i})^2E_{j,i}}{\Sigma_{j}E_{j,i}}},
\end{equation}
where $x_{j,i}$ and $E_{j,i} $ are the position of hits and the deposited energy of the $j$-th bar in the $i$-th layer, $x_{c,i}$ is the center coordinate of the shower in the $i$-th layer. It reflects the lateral development of shower, which is mainly caused by the propagation of secondary particles. Secondary particles produced by hadronic showers can propagate much farther than electromagnetic showers \cite{2020ChPhL..37k9601J}, resulting in larger values of ${\rm RMS}_{i}$. 

\begin{figure}[!htb]
\centering
\includegraphics[width=0.47\textwidth]{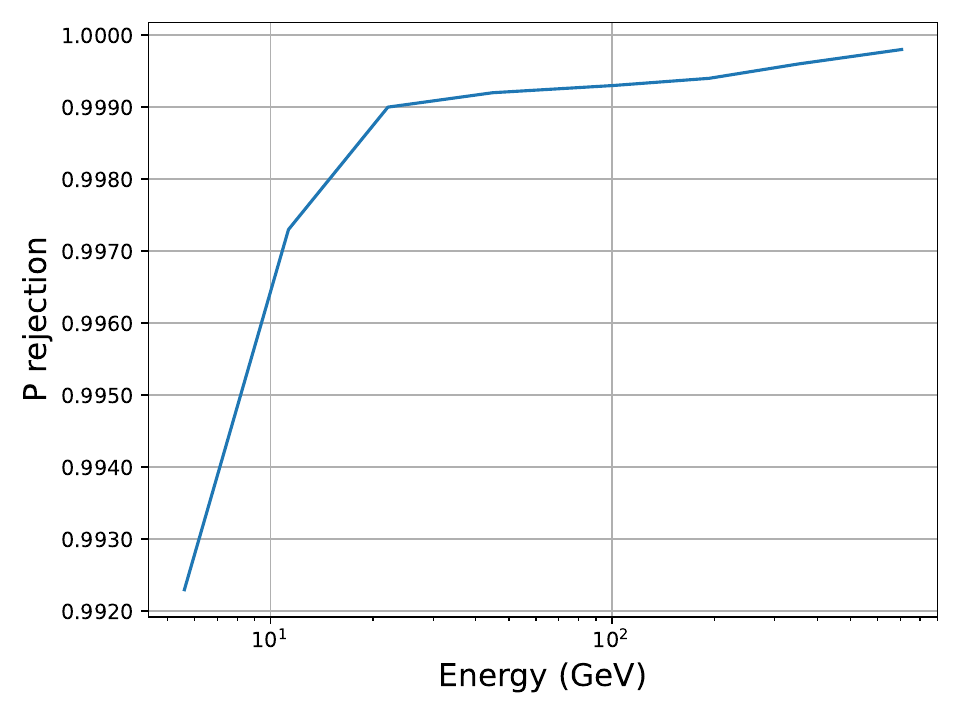}
\caption{The proton rejection fraction as a function of the energy deposition in HEIC when retaining 90\% efficiency of electrons.}
\label{fig:E_p_reject}
\end{figure}

The left panel of Fig.~\ref{fig:e-p} shows the scattering distribution of ${\rm RMS}=\sum_i {\rm RMS}_i$ and $F_{\rm last}$ for protons (orange) and electrons (blue). Both the electron and proton events have energy deposition in the HEIC of $30-100$ GeV. Electrons and protons are clearly divided into two parts. The distributions of $\zeta$ parameters of the electron and proton samples are shown in the right panel of Fig.~\ref{fig:e-p}. A $\zeta$ value of 18 is found to be able to suppress the proton contamination to $<$0.1\% on the base of maintaining a 90\% electron detection efficiency. Choosing proper cuts on the $\zeta$ parameter, we can get the rejection fraction of protons when keeping 90\% electrons, as shown in Fig. \ref{fig:E_p_reject}. At relatively low energies (several GeV), due to the small sizes of electromagnetic cascades, the proton rejection fraction is about $0.99$. With the increase of energy, the rejection fraction increases effectively, which reaches 0.999 at $\sim20$ GeV. Optimization of the e-p discrimination algorithm, e.g., by means of machine learning, can further improve the background rejection capability \cite{2022Univ....8..570X}. Combining with the background rejection fraction of $\sim 0.999$ from the ACD (design requirement) and $\sim 0.9$ from the STED based on the experience of Fermi-LAT \cite{2009ApJ...697.1071A}, the total background rejection fraction of VLAST could reach $(1-10^{-6})-(1-10^{-7})$. Thus, VLAST has an excellent ability to detect gamma rays even if the cosmic ray background flux is $10^5$ higher than photons.

VLAST is not optimized for electron detection. The vertical thickness is about 18 radiation lengths, resulting in relatively large leakages of electron events with energy above 1 TeV. However, given the large area of the detector, we can select events with large incident angles to effectively increase the slant thickness of the detector. As a reference, for incident angles $>55^{\circ}$, the thickness is about 32 radiation length, which is comparable to DAMPE. If we select events with incident angles between $55^{\circ}$ and $75^{\circ}$, the acceptance (for the HEIC-HE trigger) is about 3.5 $\rm m^2 sr$. Since DAMPE can measure the electron spectrum up to about 15 TeV, VLAST can extend the detectable upper bound of energy by at least a factor of 2, even if the spectrum is as soft as $E^{-4}$ \cite{2017Natur.552...63D}. Note, however, the dynamic range of electronics should also be expanded to detect such high energy events.

\subsection{Detectablity of gamma-ray transients}

\begin{figure}[!htb]
\centering
\includegraphics[width=0.47\textwidth]{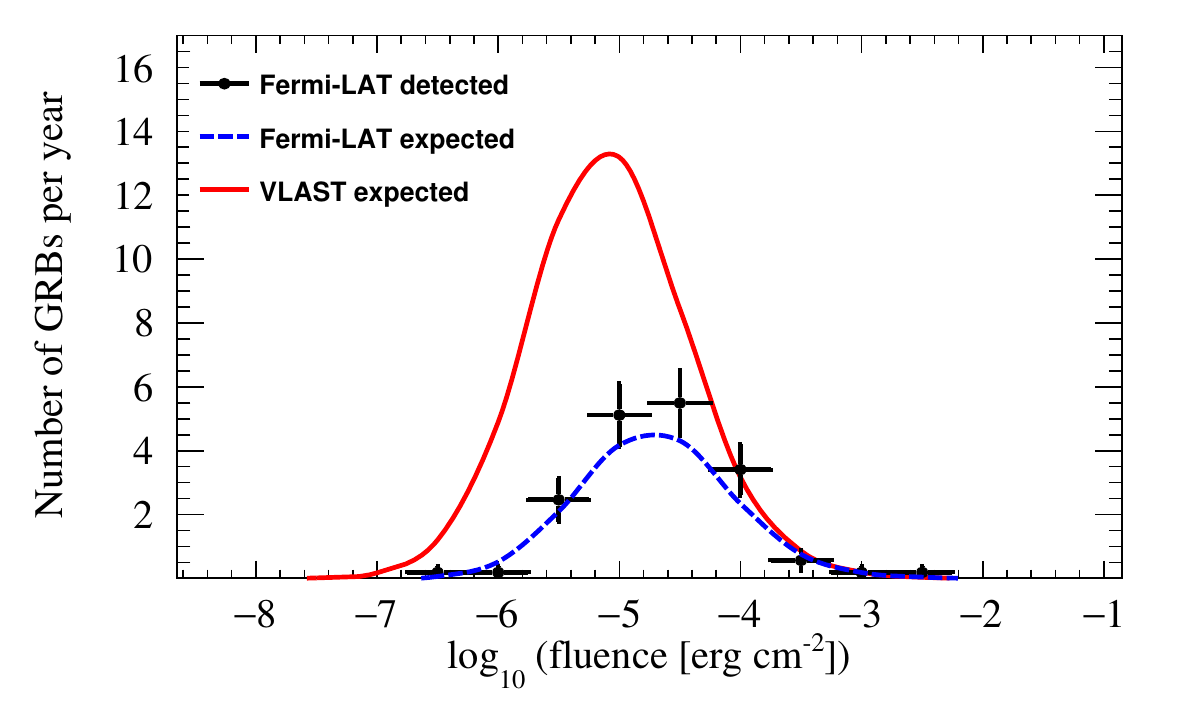}
\caption{Expected detectable rate of GRBs with different fluences by VLAST (red line), compared with the detected sample \cite{2019ApJ...878...52A} and the expectation \cite{2020ApJ...900...67K} by Fermi-LAT.}
\label{fig:GRB}
\end{figure}

VLAST has a very good capability to explore the $\gamma$-ray sky with unprecedented
sensitivity in a wide energy range \cite{2022AcASn..63...27F}. Taking the detection of burst-like
transients as an example, we briefly discuss VLAST's potential on such phenomena. Assuming a physical
sample of gamma-ray bursts (GRBs) with parameterized distributions of redshifts, luminosity, and 
spectra, one can calculate the expected detectable quantities of GRBs by a given detector with 
effective area, energy band coverage, PSF, FoV and so on 
\cite{2020ApJ...900...67K,2021ApJ...923..112X,2023ApJ...958...87Y}.
The intrinsic parameters are obtained via comparing the simulated results with the observed sample 
by various detectors. For new detectors such as VLAST, the detectability can then be calculated. 
Fig.~\ref{fig:GRB} shows the expected yearly rate for different fluences by VLAST, compared the
expected rate by Fermi-LAT \cite{2020ApJ...900...67K} and the Fermi-LAT detected rate
\cite{2019ApJ...878...52A} above 100 MeV. The total yearly rate is about 43.4 
for VLAST, which is about 3 times higher than that of Fermi-LAT (14.5 per year).

\section{Summary} \label{sec.V}
In order to validate and optimize the design of VLAST, we simulated and analyzed the performance parameters of VLAST and optimized  some details setting, such as effective area, angular resolution, energy resolution, threshold and size of ACD,width of silicon strip and so on.  VLAST directly increases the gamma-ray detection capability by increasing the effective area, while replacing the conventional tungsten plate used for electron pair conversion with CsI to detect Compton scattering events in the MeV band. This design allows the VLAST to have an effective area of 4~$\rm m^2$ which is larger than the previous ones and smaller than the APT, but with better energy and angular resolution than the APT. The outstanding performance  of VLAST can identify the uncertified point sources of Fermi-LAT, further distinguish whether the gamma-ray excess at the center of the Milky Way is due to dark matter, whether the Fermi bubble is  of leptonic or hadronic origin, and fill in the gaps in the MeV band of the extragalactic background light, and so on.  VLAST's design and validation are continuing consistently. We are considering incorporating additional sub-detectors such as time-of-flight and neutron detectors to improve  the rejection capability for complex background. The time-of-flight detectors can reduce the backsplash effect by measuring the time of recoil photons, while neutron detectors can improve the background rejection capability of the VLAST because hadron interactions produce large numbers of neutrons compared to electromagnetic interactions. The more detailed event analysis algorithms are also under development, in order to obtain more accurate performance of VLAST.  A prototype for validating the principles of the VLAST design is currently under development. The VLAST is expected to play a crucial role in gamma-ray astronomy in the future.

\acknowledgments
The authors thank Kai-Kai Duan, Yi-Qing Guo, and Yu-Hua Yao for helpful discussion.
This work is supported by the National Key Research and Development Program of China
(No. 2021YFA0718404), the National Natural Science Foundation of China (Nos. 12220101003,
12173098, U2031149), the Project for Young Scientists in Basic Research of Chinese Academy
of Sciences (CAS) (No. YSBR-061), the Scientific Instrument Developing Project of CAS
(No. GJJSTD20210009), the Youth Innovation Promotion Association of CAS, and the Young
Elite Scientists Sponsorship Program by the China Association for Science and Technology
(No. YESS20220197).

\bibliographystyle{apsrev}
\bibliography{vlast}

\begin{thebibliography}{83}
\expandafter\ifx\csname natexlab\endcsname\relax\def\natexlab#1{#1}\fi
\expandafter\ifx\csname bibnamefont\endcsname\relax
  \def\bibnamefont#1{#1}\fi
\expandafter\ifx\csname bibfnamefont\endcsname\relax
  \def\bibfnamefont#1{#1}\fi
\expandafter\ifx\csname citenamefont\endcsname\relax
  \def\citenamefont#1{#1}\fi
\expandafter\ifx\csname url\endcsname\relax
  \def\url#1{\texttt{#1}}\fi
\expandafter\ifx\csname urlprefix\endcsname\relax\def\urlprefix{URL }\fi
\providecommand{\bibinfo}[2]{#2}
\providecommand{\eprint}[2][]{\url{#2}}

\bibitem[{\citenamefont{{Kraushaar} et~al.}(1972)\citenamefont{{Kraushaar},
  {Clark}, {Garmire}, {Borken}, {Higbie}, {Leong}, and
  {Thorsos}}}]{1972ApJ...177..341K}
\bibinfo{author}{\bibfnamefont{W.~L.} \bibnamefont{{Kraushaar}}},
  \bibinfo{author}{\bibfnamefont{G.~W.} \bibnamefont{{Clark}}},
  \bibinfo{author}{\bibfnamefont{G.~P.} \bibnamefont{{Garmire}}},
  \bibinfo{author}{\bibfnamefont{R.}~\bibnamefont{{Borken}}},
  \bibinfo{author}{\bibfnamefont{P.}~\bibnamefont{{Higbie}}},
  \bibinfo{author}{\bibfnamefont{V.}~\bibnamefont{{Leong}}}, \bibnamefont{and}
  \bibinfo{author}{\bibfnamefont{T.}~\bibnamefont{{Thorsos}}},
  \bibinfo{journal}{\apj} \textbf{\bibinfo{volume}{177}}, \bibinfo{pages}{341}
  (\bibinfo{year}{1972}).

\bibitem[{\citenamefont{{Fichtel} et~al.}(1975)\citenamefont{{Fichtel},
  {Hartman}, {Kniffen}, {Thompson}, {Bignami}, {{\"O}gelman}, {{\"O}zel}, and
  {T{\"u}mer}}}]{1975ApJ...198..163F}
\bibinfo{author}{\bibfnamefont{C.~E.} \bibnamefont{{Fichtel}}},
  \bibinfo{author}{\bibfnamefont{R.~C.} \bibnamefont{{Hartman}}},
  \bibinfo{author}{\bibfnamefont{D.~A.} \bibnamefont{{Kniffen}}},
  \bibinfo{author}{\bibfnamefont{D.~J.} \bibnamefont{{Thompson}}},
  \bibinfo{author}{\bibfnamefont{G.~F.} \bibnamefont{{Bignami}}},
  \bibinfo{author}{\bibfnamefont{H.}~\bibnamefont{{{\"O}gelman}}},
  \bibinfo{author}{\bibfnamefont{M.~E.} \bibnamefont{{{\"O}zel}}},
  \bibnamefont{and}
  \bibinfo{author}{\bibfnamefont{T.}~\bibnamefont{{T{\"u}mer}}},
  \bibinfo{journal}{Astrophysical Journal} \textbf{\bibinfo{volume}{198}},
  \bibinfo{pages}{163} (\bibinfo{year}{1975}).

\bibitem[{\citenamefont{{Bignami} et~al.}(1975)\citenamefont{{Bignami},
  {Boella}, {Burger}, {Keirle}, {Mayer-Hasselwander}, {Paul}, {Pfeffermann},
  {Scarsi}, {Swanenburg}, {Taylor} et~al.}}]{1975SSI.....1..245B}
\bibinfo{author}{\bibfnamefont{G.~F.} \bibnamefont{{Bignami}}},
  \bibnamefont{et~al.}, \bibinfo{journal}{Space Science Instrumentation}
  \textbf{\bibinfo{volume}{1}}, \bibinfo{pages}{245} (\bibinfo{year}{1975}).

\bibitem[{\citenamefont{{Thompson} et~al.}(1993)\citenamefont{{Thompson},
  {Bertsch}, {Fichtel}, {Hartman}, {Hofstadter}, {Hughes}, {Hunter},
  {Hughlock}, {Kanbach}, {Kniffen} et~al.}}]{1993ApJS...86..629T}
\bibinfo{author}{\bibfnamefont{D.~J.} \bibnamefont{{Thompson}}},
  \bibnamefont{et~al.}, \bibinfo{journal}{Astrophysical Journal Supplement
  Series} \textbf{\bibinfo{volume}{86}}, \bibinfo{pages}{629}
  (\bibinfo{year}{1993}).

\bibitem[{\citenamefont{{Atwood} et~al.}(2009)\citenamefont{{Atwood}, {Abdo},
  {Ackermann}, {Althouse}, {Anderson}, {Axelsson}, {Baldini}, {Ballet}, {Band},
  {Barbiellini} et~al.}}]{2009ApJ...697.1071A}
\bibinfo{author}{\bibfnamefont{W.~B.} \bibnamefont{{Atwood}}},
  \bibnamefont{et~al.}, \bibinfo{journal}{Astrophysical Journal}
  \textbf{\bibinfo{volume}{697}}, \bibinfo{pages}{1071} (\bibinfo{year}{2009}),
  \eprint{0902.1089}.

\bibitem[{\citenamefont{{Abdollahi} et~al.}(2022)\citenamefont{{Abdollahi},
  {Acero}, {Baldini}, {Ballet}, {Bastieri}, {Bellazzini}, {Berenji},
  {Berretta}, {Bissaldi}, {Blandford} et~al.}}]{2022ApJS..260...53A}
\bibinfo{author}{\bibfnamefont{S.}~\bibnamefont{{Abdollahi}}},
  \bibnamefont{et~al.}, \bibinfo{journal}{Astrophysical Journal Supplement
  Series} \textbf{\bibinfo{volume}{260}}, \bibinfo{eid}{53}
  (\bibinfo{year}{2022}), \eprint{2201.11184}.

\bibitem[{\citenamefont{{Topchiev} et~al.}(2022)\citenamefont{{Topchiev},
  {Galper}, {Arkhangelskaja}, {Arkhangelskiy}, {Bakaldin}, {Cherniy},
  {Chernysheva}, {Gudkova}, {Gusakov}, {Dalkarov}
  et~al.}}]{2022AdSpR..70.2773T}
\bibinfo{author}{\bibfnamefont{N.~P.} \bibnamefont{{Topchiev}}},
  \bibnamefont{et~al.}, \bibinfo{journal}{Advances in Space Research}
  \textbf{\bibinfo{volume}{70}}, \bibinfo{pages}{2773} (\bibinfo{year}{2022}),
  \eprint{2108.12609}.

\bibitem[{\citenamefont{{Schoenfelder}
  et~al.}(1993)\citenamefont{{Schoenfelder}, {Aarts}, {Bennett}, {de Boer},
  {Clear}, {Collmar}, {Connors}, {Deerenberg}, {Diehl}, {von Dordrecht}
  et~al.}}]{1993ApJS...86..657S}
\bibinfo{author}{\bibfnamefont{V.}~\bibnamefont{{Schoenfelder}}},
  \bibnamefont{et~al.}, \bibinfo{journal}{Astrophysical Journal Supplement
  Series} \textbf{\bibinfo{volume}{86}}, \bibinfo{pages}{657}
  (\bibinfo{year}{1993}).

\bibitem[{\citenamefont{{Kole} et~al.}(2020)\citenamefont{{Kole}, {De Angelis},
  {Berlato}, {Burgess}, {Gauvin}, {Greiner}, {Hajdas}, {Li}, {Li}, {Pollo}
  et~al.}}]{2020A&A...644A.124K}
\bibinfo{author}{\bibfnamefont{M.}~\bibnamefont{{Kole}}}, \bibnamefont{et~al.},
  \bibinfo{journal}{Astronomy \& Astrophysics} \textbf{\bibinfo{volume}{644}},
  \bibinfo{eid}{A124} (\bibinfo{year}{2020}), \eprint{2009.04871}.

\bibitem[{\citenamefont{{Zoglauer} et~al.}(2021)\citenamefont{{Zoglauer},
  {Siegert}, {Lowell}, {Mochizuki}, {Kierans}, {Sleator}, {Hartmann}, {Lazar},
  {Gulick}, {Beechert} et~al.}}]{2021arXiv210213158Z}
\bibinfo{author}{\bibfnamefont{A.}~\bibnamefont{{Zoglauer}}},
  \bibnamefont{et~al.}, \bibinfo{journal}{arXiv e-prints}
  \bibinfo{eid}{arXiv:2102.13158} (\bibinfo{year}{2021}), \eprint{2102.13158}.

\bibitem[{\citenamefont{{Wu} et~al.}(2014)\citenamefont{{Wu}, {Su}, {Bravar},
  {Chang}, {Fan}, {Pohl}, and {Walter}}}]{2014SPIE.9144E..0FW}
\bibinfo{author}{\bibfnamefont{X.}~\bibnamefont{{Wu}}},
  \bibinfo{author}{\bibfnamefont{M.}~\bibnamefont{{Su}}},
  \bibinfo{author}{\bibfnamefont{A.}~\bibnamefont{{Bravar}}},
  \bibinfo{author}{\bibfnamefont{J.}~\bibnamefont{{Chang}}},
  \bibinfo{author}{\bibfnamefont{Y.}~\bibnamefont{{Fan}}},
  \bibinfo{author}{\bibfnamefont{M.}~\bibnamefont{{Pohl}}}, \bibnamefont{and}
  \bibinfo{author}{\bibfnamefont{R.}~\bibnamefont{{Walter}}}, in
  \emph{\bibinfo{booktitle}{Space Telescopes and Instrumentation 2014:
  Ultraviolet to Gamma Ray}}, edited by
  \bibinfo{editor}{\bibfnamefont{T.}~\bibnamefont{{Takahashi}}},
  \bibinfo{editor}{\bibfnamefont{J.-W.~A.} \bibnamefont{{den Herder}}},
  \bibnamefont{and} \bibinfo{editor}{\bibfnamefont{M.}~\bibnamefont{{Bautz}}}
  (\bibinfo{year}{2014}), vol. \bibinfo{volume}{9144} of
  \emph{\bibinfo{series}{Society of Photo-Optical Instrumentation Engineers
  (SPIE) Conference Series}}, p. \bibinfo{pages}{91440F}, \eprint{1407.0710}.

\bibitem[{\citenamefont{{McEnery} et~al.}(2019)\citenamefont{{McEnery}, {van
  der Horst}, {Dominguez}, {Moiseev}, {Marcowith}, {Harding}, {Lien},
  {Giuliani}, {Inglis}, {Ansoldi} et~al.}}]{2019BAAS...51g.245M}
\bibinfo{author}{\bibfnamefont{J.}~\bibnamefont{{McEnery}}},
  \bibnamefont{et~al.}, in \emph{\bibinfo{booktitle}{Bulletin of the American
  Astronomical Society}} (\bibinfo{year}{2019}), vol.~\bibinfo{volume}{51}, p.
  \bibinfo{pages}{245}, \eprint{1907.07558}.

\bibitem[{\citenamefont{{De Angelis} et~al.}(2017)\citenamefont{{De Angelis},
  {Tatischeff}, {Tavani}, {Oberlack}, {Grenier}, {Hanlon}, {Walter}, {Argan},
  {von Ballmoos}, {Bulgarelli} et~al.}}]{2017ExA....44...25D}
\bibinfo{author}{\bibfnamefont{A.}~\bibnamefont{{De Angelis}}},
  \bibnamefont{et~al.}, \bibinfo{journal}{Experimental Astronomy}
  \textbf{\bibinfo{volume}{44}}, \bibinfo{pages}{25} (\bibinfo{year}{2017}),
  \eprint{1611.02232}.

\bibitem[{\citenamefont{{Hunter} et~al.}(2014)\citenamefont{{Hunter}, {Bloser},
  {Depaola}, {Dion}, {DeNolfo}, {Hanu}, {Iparraguirre}, {Legere}, {Longo},
  {McConnell} et~al.}}]{2014APh....59...18H}
\bibinfo{author}{\bibfnamefont{S.~D.} \bibnamefont{{Hunter}}},
  \bibnamefont{et~al.}, \bibinfo{journal}{Astroparticle Physics}
  \textbf{\bibinfo{volume}{59}}, \bibinfo{pages}{18} (\bibinfo{year}{2014}),
  \eprint{1311.2059}.

\bibitem[{\citenamefont{{Orlando} et~al.}(2022)\citenamefont{{Orlando},
  {Bottacini}, {Moiseev}, {Bodaghee}, {Collmar}, {Ensslin}, {Moskalenko},
  {Negro}, {Profumo}, {Digel} et~al.}}]{2022JCAP...07..036O}
\bibinfo{author}{\bibfnamefont{E.}~\bibnamefont{{Orlando}}},
  \bibnamefont{et~al.}, \bibinfo{journal}{Journal of Cosmology and
  Astroparticle Physics} \textbf{\bibinfo{volume}{2022}}, \bibinfo{eid}{036}
  (\bibinfo{year}{2022}), \eprint{2112.07190}.

\bibitem[{\citenamefont{{Dzhatdoev} and
  {Podlesnyi}}(2019)}]{2019APh...112....1D}
\bibinfo{author}{\bibfnamefont{T.}~\bibnamefont{{Dzhatdoev}}} \bibnamefont{and}
  \bibinfo{author}{\bibfnamefont{E.}~\bibnamefont{{Podlesnyi}}},
  \bibinfo{journal}{Astroparticle Physics} \textbf{\bibinfo{volume}{112}},
  \bibinfo{pages}{1} (\bibinfo{year}{2019}), \eprint{1902.01491}.

\bibitem[{\citenamefont{{Aramaki} et~al.}(2020)\citenamefont{{Aramaki},
  {Adrian}, {Karagiorgi}, and {Odaka}}}]{2020APh...114..107A}
\bibinfo{author}{\bibfnamefont{T.}~\bibnamefont{{Aramaki}}},
  \bibinfo{author}{\bibfnamefont{P.~O.~H.} \bibnamefont{{Adrian}}},
  \bibinfo{author}{\bibfnamefont{G.}~\bibnamefont{{Karagiorgi}}},
  \bibnamefont{and} \bibinfo{author}{\bibfnamefont{H.}~\bibnamefont{{Odaka}}},
  \bibinfo{journal}{Astroparticle Physics} \textbf{\bibinfo{volume}{114}},
  \bibinfo{pages}{107} (\bibinfo{year}{2020}), \eprint{1901.03430}.

\bibitem[{\citenamefont{{Labanti} et~al.}(2021)\citenamefont{{Labanti},
  {Amati}, {Frontera}, {Mereghetti}, {Gasent-Blesa}, {Tenzer}, {Orleanski},
  {Kuvvetli}, {Campana}, {Fuschino} et~al.}}]{2021arXiv210208701L}
\bibinfo{author}{\bibfnamefont{C.}~\bibnamefont{{Labanti}}},
  \bibnamefont{et~al.}, \bibinfo{journal}{arXiv e-prints}
  \bibinfo{eid}{arXiv:2102.08701} (\bibinfo{year}{2021}), \eprint{2102.08701}.

\bibitem[{\citenamefont{{Barbato} et~al.}(2022)\citenamefont{{Barbato}, {Abba},
  {Anastasio}, {Barbarino}, {Boiano}, {de Asmundis}, {De Mitri}, {Ferrentino},
  {Garufi}, {Guarino} et~al.}}]{2022icrc.confE.581B}
\bibinfo{author}{\bibfnamefont{F.}~\bibnamefont{{Barbato}}},
  \bibnamefont{et~al.}, in \emph{\bibinfo{booktitle}{37th International Cosmic
  Ray Conference}} (\bibinfo{year}{2022}), p. \bibinfo{pages}{581}.

\bibitem[{\citenamefont{{Zhu} et~al.}(2023)\citenamefont{{Zhu}, {Zheng},
  {Feng}, {Zeng}, {Huang}, {Hsiang}, {Chang}, {Li}, {Chang}, {Pan}
  et~al.}}]{2023arXiv231211900Z}
\bibinfo{author}{\bibfnamefont{J.}~\bibnamefont{{Zhu}}}, \bibnamefont{et~al.},
  \bibinfo{journal}{arXiv e-prints} \bibinfo{eid}{arXiv:2312.11900}
  (\bibinfo{year}{2023}), \eprint{2312.11900}.

\bibitem[{\citenamefont{{Fan} et~al.}(2022)\citenamefont{{Fan}, {Chang}, {Guo},
  {Yuan}, {Hu}, {Li}, {Yue}, {Huang}, {Liu}, {Feng}
  et~al.}}]{2022AcASn..63...27F}
\bibinfo{author}{\bibfnamefont{Y.~Z.} \bibnamefont{{Fan}}},
  \bibnamefont{et~al.}, \bibinfo{journal}{Acta Astronomica Sinica}
  \textbf{\bibinfo{volume}{63}}, \bibinfo{eid}{27} (\bibinfo{year}{2022}).

\bibitem[{\citenamefont{{Wang} et~al.}(2022)\citenamefont{{Wang}, {Feng},
  {Luo}, {Zhang}, and {Liu}}}]{2022cosp...44.3058W}
\bibinfo{author}{\bibfnamefont{Y.}~\bibnamefont{{Wang}}},
  \bibinfo{author}{\bibfnamefont{C.}~\bibnamefont{{Feng}}},
  \bibinfo{author}{\bibfnamefont{L.}~\bibnamefont{{Luo}}},
  \bibinfo{author}{\bibfnamefont{Y.}~\bibnamefont{{Zhang}}}, \bibnamefont{and}
  \bibinfo{author}{\bibfnamefont{S.}~\bibnamefont{{Liu}}}, in
  \emph{\bibinfo{booktitle}{44th COSPAR Scientific Assembly. Held 16-24 July}}
  (\bibinfo{year}{2022}), vol.~\bibinfo{volume}{44}, p. \bibinfo{pages}{3058}.

\bibitem[{\citenamefont{Wan et~al.}(2023)\citenamefont{Wan, Guo, Xu, Wang,
  Zhang, Hu, Zhang, Pan, Li, Yue et~al.}}]{wan2023design}
\bibinfo{author}{\bibfnamefont{Q.}~\bibnamefont{Wan}}, \bibnamefont{et~al.},
  \bibinfo{journal}{Nuclear Science and Techniques}
  \textbf{\bibinfo{volume}{34}}, \bibinfo{pages}{149} (\bibinfo{year}{2023}).

\bibitem[{\citenamefont{Yang et~al.}(2022)\citenamefont{Yang, Li, Yu, Chen,
  Kong, Zhang, Tang, Guo, Yang, and Su}}]{Yang:2022eaq}
\bibinfo{author}{\bibfnamefont{H.-B.} \bibnamefont{Yang}},
  \bibinfo{author}{\bibfnamefont{X.-Q.} \bibnamefont{Li}},
  \bibinfo{author}{\bibfnamefont{Y.-H.} \bibnamefont{Yu}},
  \bibinfo{author}{\bibfnamefont{Y.}~\bibnamefont{Chen}},
  \bibinfo{author}{\bibfnamefont{J.}~\bibnamefont{Kong}},
  \bibinfo{author}{\bibfnamefont{Y.-J.} \bibnamefont{Zhang}},
  \bibinfo{author}{\bibfnamefont{S.-W.} \bibnamefont{Tang}},
  \bibinfo{author}{\bibfnamefont{J.-H.} \bibnamefont{Guo}},
  \bibinfo{author}{\bibfnamefont{B.}~\bibnamefont{Yang}}, \bibnamefont{and}
  \bibinfo{author}{\bibfnamefont{F.-J.} \bibnamefont{Su}},
  \bibinfo{journal}{Nuclear Science and Techniques}
  \textbf{\bibinfo{volume}{33}}, \bibinfo{pages}{65} (\bibinfo{year}{2022}).

\bibitem[{\citenamefont{{Chang} et~al.}(2017)\citenamefont{{Chang}, {Ambrosi},
  {An}, {Asfandiyarov}, {Azzarello}, {Bernardini}, {Bertucci}, {Cai},
  {Caragiulo}, {Chen} et~al.}}]{2017APh....95....6C}
\bibinfo{author}{\bibfnamefont{J.}~\bibnamefont{{Chang}}},
  \bibnamefont{et~al.}, \bibinfo{journal}{Astroparticle Physics}
  \textbf{\bibinfo{volume}{95}}, \bibinfo{pages}{6} (\bibinfo{year}{2017}),
  \eprint{1706.08453}.

\bibitem[{\citenamefont{{Buckley} et~al.}(2022)\citenamefont{{Buckley},
  {Adapt}, {Alnussirat}, {Altomare}, {Bose}, {Braun}, {Buckley}, {Buhler},
  {Burns}, {Chamberlain} et~al.}}]{2022icrc.confE.655B}
\bibinfo{author}{\bibfnamefont{J.}~\bibnamefont{{Buckley}}},
  \bibnamefont{et~al.}, in \emph{\bibinfo{booktitle}{37th International Cosmic
  Ray Conference}} (\bibinfo{year}{2022}), p. \bibinfo{pages}{655}.

\bibitem[{\citenamefont{{Xia} et~al.}(2018)\citenamefont{{Xia}, {Zhang},
  {Liang}, {Feng}, {Yuan}, {Fan}, and {Wu}}}]{2018PhRvD..97f3003X}
\bibinfo{author}{\bibfnamefont{Z.-Q.} \bibnamefont{{Xia}}},
  \bibinfo{author}{\bibfnamefont{C.}~\bibnamefont{{Zhang}}},
  \bibinfo{author}{\bibfnamefont{Y.-F.} \bibnamefont{{Liang}}},
  \bibinfo{author}{\bibfnamefont{L.}~\bibnamefont{{Feng}}},
  \bibinfo{author}{\bibfnamefont{Q.}~\bibnamefont{{Yuan}}},
  \bibinfo{author}{\bibfnamefont{Y.-Z.} \bibnamefont{{Fan}}}, \bibnamefont{and}
  \bibinfo{author}{\bibfnamefont{J.}~\bibnamefont{{Wu}}},
  \bibinfo{journal}{Physical Review D} \textbf{\bibinfo{volume}{97}},
  \bibinfo{eid}{063003} (\bibinfo{year}{2018}), \eprint{1801.01646}.

\bibitem[{\citenamefont{{Liang} et~al.}(2016)\citenamefont{{Liang}, {Shen},
  {Li}, {Fan}, {Huang}, {Lei}, {Feng}, {Liang}, and
  {Chang}}}]{2016PhRvD..93j3525L}
\bibinfo{author}{\bibfnamefont{Y.-F.} \bibnamefont{{Liang}}},
  \bibinfo{author}{\bibfnamefont{Z.-Q.} \bibnamefont{{Shen}}},
  \bibinfo{author}{\bibfnamefont{X.}~\bibnamefont{{Li}}},
  \bibinfo{author}{\bibfnamefont{Y.-Z.} \bibnamefont{{Fan}}},
  \bibinfo{author}{\bibfnamefont{X.}~\bibnamefont{{Huang}}},
  \bibinfo{author}{\bibfnamefont{S.-J.} \bibnamefont{{Lei}}},
  \bibinfo{author}{\bibfnamefont{L.}~\bibnamefont{{Feng}}},
  \bibinfo{author}{\bibfnamefont{E.-W.} \bibnamefont{{Liang}}},
  \bibnamefont{and} \bibinfo{author}{\bibfnamefont{J.}~\bibnamefont{{Chang}}},
  \bibinfo{journal}{Physical Review D} \textbf{\bibinfo{volume}{93}},
  \bibinfo{eid}{103525} (\bibinfo{year}{2016}), \eprint{1602.06527}.

\bibitem[{\citenamefont{{Bringmann} et~al.}(2012)\citenamefont{{Bringmann},
  {Huang}, {Ibarra}, {Vogl}, and {Weniger}}}]{2012JCAP...07..054B}
\bibinfo{author}{\bibfnamefont{T.}~\bibnamefont{{Bringmann}}},
  \bibinfo{author}{\bibfnamefont{X.}~\bibnamefont{{Huang}}},
  \bibinfo{author}{\bibfnamefont{A.}~\bibnamefont{{Ibarra}}},
  \bibinfo{author}{\bibfnamefont{S.}~\bibnamefont{{Vogl}}}, \bibnamefont{and}
  \bibinfo{author}{\bibfnamefont{C.}~\bibnamefont{{Weniger}}},
  \bibinfo{journal}{Journal of Cosmology and Astroparticle Physics}
  \textbf{\bibinfo{volume}{2012}}, \bibinfo{eid}{054} (\bibinfo{year}{2012}),
  \eprint{1203.1312}.

\bibitem[{\citenamefont{{Wang} and {Han}}(2013)}]{2013PhRvD..87a5015W}
\bibinfo{author}{\bibfnamefont{L.}~\bibnamefont{{Wang}}} \bibnamefont{and}
  \bibinfo{author}{\bibfnamefont{X.-F.} \bibnamefont{{Han}}},
  \bibinfo{journal}{\prd} \textbf{\bibinfo{volume}{87}}, \bibinfo{eid}{015015}
  (\bibinfo{year}{2013}), \eprint{1209.0376}.

\bibitem[{\citenamefont{{Zhou} et~al.}(2015)\citenamefont{{Zhou}, {Liang},
  {Huang}, {Li}, {Fan}, {Feng}, and {Chang}}}]{2015PhRvD..91l3010Z}
\bibinfo{author}{\bibfnamefont{B.}~\bibnamefont{{Zhou}}},
  \bibinfo{author}{\bibfnamefont{Y.-F.} \bibnamefont{{Liang}}},
  \bibinfo{author}{\bibfnamefont{X.}~\bibnamefont{{Huang}}},
  \bibinfo{author}{\bibfnamefont{X.}~\bibnamefont{{Li}}},
  \bibinfo{author}{\bibfnamefont{Y.-Z.} \bibnamefont{{Fan}}},
  \bibinfo{author}{\bibfnamefont{L.}~\bibnamefont{{Feng}}}, \bibnamefont{and}
  \bibinfo{author}{\bibfnamefont{J.}~\bibnamefont{{Chang}}},
  \bibinfo{journal}{\prd} \textbf{\bibinfo{volume}{91}}, \bibinfo{eid}{123010}
  (\bibinfo{year}{2015}), \eprint{1406.6948}.

\bibitem[{\citenamefont{{Su} et~al.}(2010)\citenamefont{{Su}, {Slatyer}, and
  {Finkbeiner}}}]{2010ApJ...724.1044S}
\bibinfo{author}{\bibfnamefont{M.}~\bibnamefont{{Su}}},
  \bibinfo{author}{\bibfnamefont{T.~R.} \bibnamefont{{Slatyer}}},
  \bibnamefont{and} \bibinfo{author}{\bibfnamefont{D.~P.}
  \bibnamefont{{Finkbeiner}}}, \bibinfo{journal}{\apj}
  \textbf{\bibinfo{volume}{724}}, \bibinfo{pages}{1044} (\bibinfo{year}{2010}),
  \eprint{1005.5480}.

\bibitem[{\citenamefont{{Zhang}}(2019)}]{2019Natur.575..448Z}
\bibinfo{author}{\bibfnamefont{B.}~\bibnamefont{{Zhang}}},
  \bibinfo{journal}{Nature} \textbf{\bibinfo{volume}{575}},
  \bibinfo{pages}{448} (\bibinfo{year}{2019}), \eprint{1911.09862}.

\bibitem[{\citenamefont{{Xing} and {Wang}}(2016)}]{2016ApJ...831..143X}
\bibinfo{author}{\bibfnamefont{Y.}~\bibnamefont{{Xing}}} \bibnamefont{and}
  \bibinfo{author}{\bibfnamefont{Z.}~\bibnamefont{{Wang}}},
  \bibinfo{journal}{\apj} \textbf{\bibinfo{volume}{831}}, \bibinfo{eid}{143}
  (\bibinfo{year}{2016}), \eprint{1604.08710}.

\bibitem[{\citenamefont{{Yuan} and {Zhang}}(2014)}]{2014JHEAp...3....1Y}
\bibinfo{author}{\bibfnamefont{Q.}~\bibnamefont{{Yuan}}} \bibnamefont{and}
  \bibinfo{author}{\bibfnamefont{B.}~\bibnamefont{{Zhang}}},
  \bibinfo{journal}{Journal of High Energy Astrophysics}
  \textbf{\bibinfo{volume}{3}}, \bibinfo{pages}{1} (\bibinfo{year}{2014}),
  \eprint{1404.2318}.

\bibitem[{\citenamefont{{Xing} et~al.}(2016)\citenamefont{{Xing}, {Wang},
  {Zhang}, and {Chen}}}]{2016ApJ...823...44X}
\bibinfo{author}{\bibfnamefont{Y.}~\bibnamefont{{Xing}}},
  \bibinfo{author}{\bibfnamefont{Z.}~\bibnamefont{{Wang}}},
  \bibinfo{author}{\bibfnamefont{X.}~\bibnamefont{{Zhang}}}, \bibnamefont{and}
  \bibinfo{author}{\bibfnamefont{Y.}~\bibnamefont{{Chen}}},
  \bibinfo{journal}{\apj} \textbf{\bibinfo{volume}{823}}, \bibinfo{eid}{44}
  (\bibinfo{year}{2016}), \eprint{1603.00998}.

\bibitem[{\citenamefont{{Liu} et~al.}(2015)\citenamefont{{Liu}, {Chen},
  {Zhang}, {Zhang}, {Xing}, and {Pannuti}}}]{2015ApJ...809..102L}
\bibinfo{author}{\bibfnamefont{B.}~\bibnamefont{{Liu}}},
  \bibinfo{author}{\bibfnamefont{Y.}~\bibnamefont{{Chen}}},
  \bibinfo{author}{\bibfnamefont{X.}~\bibnamefont{{Zhang}}},
  \bibinfo{author}{\bibfnamefont{G.-Y.} \bibnamefont{{Zhang}}},
  \bibinfo{author}{\bibfnamefont{Y.}~\bibnamefont{{Xing}}}, \bibnamefont{and}
  \bibinfo{author}{\bibfnamefont{T.~G.} \bibnamefont{{Pannuti}}},
  \bibinfo{journal}{\apj} \textbf{\bibinfo{volume}{809}}, \bibinfo{eid}{102}
  (\bibinfo{year}{2015}), \eprint{1502.02679}.

\bibitem[{\citenamefont{{Acero} et~al.}(2016)\citenamefont{{Acero},
  {Ackermann}, {Ajello}, {Baldini}, {Ballet}, {Barbiellini}, {Bastieri},
  {Bellazzini}, {Bissaldi}, {Blandford} et~al.}}]{2016ApJS..224....8A}
\bibinfo{author}{\bibfnamefont{F.}~\bibnamefont{{Acero}}},
  \bibnamefont{et~al.}, \bibinfo{journal}{Astrophysical Journal Supplement
  Series} \textbf{\bibinfo{volume}{224}}, \bibinfo{eid}{8}
  (\bibinfo{year}{2016}), \eprint{1511.06778}.

\bibitem[{\citenamefont{{Ackermann}
  et~al.}(2013{\natexlab{a}})\citenamefont{{Ackermann}, {Ajello}, {Allafort},
  {Baldini}, {Ballet}, {Barbiellini}, {Baring}, {Bastieri}, {Bechtol},
  {Bellazzini} et~al.}}]{2013Sci...339..807A}
\bibinfo{author}{\bibfnamefont{M.}~\bibnamefont{{Ackermann}}},
  \bibnamefont{et~al.}, \bibinfo{journal}{Science}
  \textbf{\bibinfo{volume}{339}}, \bibinfo{pages}{807}
  (\bibinfo{year}{2013}{\natexlab{a}}), \eprint{1302.3307}.

\bibitem[{\citenamefont{{Yuan} et~al.}(2017)\citenamefont{{Yuan}, {Lin},
  {Fang}, and {Bi}}}]{2017PhRvD..95h3007Y}
\bibinfo{author}{\bibfnamefont{Q.}~\bibnamefont{{Yuan}}},
  \bibinfo{author}{\bibfnamefont{S.-J.} \bibnamefont{{Lin}}},
  \bibinfo{author}{\bibfnamefont{K.}~\bibnamefont{{Fang}}}, \bibnamefont{and}
  \bibinfo{author}{\bibfnamefont{X.-J.} \bibnamefont{{Bi}}},
  \bibinfo{journal}{\prd} \textbf{\bibinfo{volume}{95}}, \bibinfo{eid}{083007}
  (\bibinfo{year}{2017}), \eprint{1701.06149}.

\bibitem[{\citenamefont{{Chen} et~al.}(2022)\citenamefont{{Chen}, {Zhang}, and
  {Long}}}]{2022PhRvD.105f3008C}
\bibinfo{author}{\bibfnamefont{S.}~\bibnamefont{{Chen}}},
  \bibinfo{author}{\bibfnamefont{H.-H.} \bibnamefont{{Zhang}}},
  \bibnamefont{and} \bibinfo{author}{\bibfnamefont{G.}~\bibnamefont{{Long}}},
  \bibinfo{journal}{\prd} \textbf{\bibinfo{volume}{105}}, \bibinfo{eid}{063008}
  (\bibinfo{year}{2022}), \eprint{2112.15463}.

\bibitem[{\citenamefont{{Roth} et~al.}(2021)\citenamefont{{Roth}, {Krumholz},
  {Crocker}, and {Celli}}}]{2021Natur.597..341R}
\bibinfo{author}{\bibfnamefont{M.~A.} \bibnamefont{{Roth}}},
  \bibinfo{author}{\bibfnamefont{M.~R.} \bibnamefont{{Krumholz}}},
  \bibinfo{author}{\bibfnamefont{R.~M.} \bibnamefont{{Crocker}}},
  \bibnamefont{and} \bibinfo{author}{\bibfnamefont{S.}~\bibnamefont{{Celli}}},
  \bibinfo{journal}{Nature} \textbf{\bibinfo{volume}{597}},
  \bibinfo{pages}{341} (\bibinfo{year}{2021}), \eprint{2109.07598}.

\bibitem[{\citenamefont{{Fermi-LAT Collaboration}
  et~al.}(2018)\citenamefont{{Fermi-LAT Collaboration}, {Abdollahi},
  {Ackermann}, {Ajello}, {Atwood}, {Baldini}, {Ballet}, {Barbiellini},
  {Bastieri}, {Becerra Gonzalez} et~al.}}]{2018Sci...362.1031F}
\bibinfo{author}{\bibnamefont{{Fermi-LAT Collaboration}}},
  \bibnamefont{et~al.}, \bibinfo{journal}{Science}
  \textbf{\bibinfo{volume}{362}}, \bibinfo{pages}{1031} (\bibinfo{year}{2018}),
  \eprint{1812.01031}.

\bibitem[{\citenamefont{{Zeng} and {Yan}}(2019)}]{2019ApJ...882...87Z}
\bibinfo{author}{\bibfnamefont{H.}~\bibnamefont{{Zeng}}} \bibnamefont{and}
  \bibinfo{author}{\bibfnamefont{D.}~\bibnamefont{{Yan}}},
  \bibinfo{journal}{\apj} \textbf{\bibinfo{volume}{882}}, \bibinfo{eid}{87}
  (\bibinfo{year}{2019}), \eprint{1907.10965}.

\bibitem[{\citenamefont{{Wei} and {Wu}}(2021)}]{2021FrPhy..1644300W}
\bibinfo{author}{\bibfnamefont{J.-J.} \bibnamefont{{Wei}}} \bibnamefont{and}
  \bibinfo{author}{\bibfnamefont{X.-F.} \bibnamefont{{Wu}}},
  \bibinfo{journal}{Frontiers of Physics} \textbf{\bibinfo{volume}{16}},
  \bibinfo{eid}{44300} (\bibinfo{year}{2021}), \eprint{2102.03724}.

\bibitem[{\citenamefont{{Amelino-Camelia}
  et~al.}(1998)\citenamefont{{Amelino-Camelia}, {Ellis}, {Mavromatos},
  {Nanopoulos}, and {Sarkar}}}]{1998Natur.393..763A}
\bibinfo{author}{\bibfnamefont{G.}~\bibnamefont{{Amelino-Camelia}}},
  \bibinfo{author}{\bibfnamefont{J.}~\bibnamefont{{Ellis}}},
  \bibinfo{author}{\bibfnamefont{N.~E.} \bibnamefont{{Mavromatos}}},
  \bibinfo{author}{\bibfnamefont{D.~V.} \bibnamefont{{Nanopoulos}}},
  \bibnamefont{and} \bibinfo{author}{\bibfnamefont{S.}~\bibnamefont{{Sarkar}}},
  \bibinfo{journal}{Nature} \textbf{\bibinfo{volume}{393}},
  \bibinfo{pages}{763} (\bibinfo{year}{1998}), \eprint{astro-ph/9712103}.

\bibitem[{\citenamefont{{Li} and {Ma}}(2023)}]{2023APh...14802831L}
\bibinfo{author}{\bibfnamefont{H.}~\bibnamefont{{Li}}} \bibnamefont{and}
  \bibinfo{author}{\bibfnamefont{B.-Q.} \bibnamefont{{Ma}}},
  \bibinfo{journal}{Astroparticle Physics} \textbf{\bibinfo{volume}{148}},
  \bibinfo{eid}{102831} (\bibinfo{year}{2023}), \eprint{2210.06338}.

\bibitem[{\citenamefont{{Cao} et~al.}(2022)\citenamefont{{Cao}, {Aharonian},
  {An}, {Axikegu}, {Bai}, {Bao}, {Bastieri}, {Bi}, {Bi}, {Cai}
  et~al.}}]{2022PhRvL.128e1102C}
\bibinfo{author}{\bibfnamefont{Z.}~\bibnamefont{{Cao}}}, \bibnamefont{et~al.},
  \bibinfo{journal}{\prl} \textbf{\bibinfo{volume}{128}}, \bibinfo{eid}{051102}
  (\bibinfo{year}{2022}).

\bibitem[{\citenamefont{{Allison} et~al.}(2016)\citenamefont{{Allison},
  {Amako}, {Apostolakis}, {Arce}, {Asai}, {Aso}, {Bagli}, {Bagulya},
  {Banerjee}, {Barrand} et~al.}}]{2016NIMPA.835..186A}
\bibinfo{author}{\bibfnamefont{J.}~\bibnamefont{{Allison}}},
  \bibnamefont{et~al.}, \bibinfo{journal}{Nuclear Instruments and Methods in
  Physics Research A} \textbf{\bibinfo{volume}{835}}, \bibinfo{pages}{186}
  (\bibinfo{year}{2016}).

\bibitem[{\citenamefont{{Tavani} et~al.}(2009)\citenamefont{{Tavani},
  {Barbiellini}, {Argan}, {Boffelli}, {Bulgarelli}, {Caraveo}, {Cattaneo},
  {Chen}, {Cocco}, {Costa} et~al.}}]{2009A&A...502..995T}
\bibinfo{author}{\bibfnamefont{M.}~\bibnamefont{{Tavani}}},
  \bibnamefont{et~al.}, \bibinfo{journal}{Astronomy \& Astrophysics}
  \textbf{\bibinfo{volume}{502}}, \bibinfo{pages}{995} (\bibinfo{year}{2009}),
  \eprint{0807.4254}.

\bibitem[{\citenamefont{{Allison} et~al.}(2006)\citenamefont{{Allison},
  {Amako}, {Apostolakis}, {Araujo}, {Dubois}, {Asai}, {Barrand}, {Capra},
  {Chauvie}, {Chytracek} et~al.}}]{2006ITNS...53..270A}
\bibinfo{author}{\bibfnamefont{J.}~\bibnamefont{{Allison}}},
  \bibnamefont{et~al.}, \bibinfo{journal}{IEEE Transactions on Nuclear Science}
  \textbf{\bibinfo{volume}{53}}, \bibinfo{pages}{270} (\bibinfo{year}{2006}).

\bibitem[{\citenamefont{{Aad} et~al.}(2010)\citenamefont{{Aad}, {Abbott},
  {Abdallah}, {Abdelalim}, {Abdesselam}, {Abdinov}, {Abi}, {Abolins},
  {Abramowicz}, {Abreu} et~al.}}]{2010EPJC...70..823A}
\bibinfo{author}{\bibfnamefont{G.}~\bibnamefont{{Aad}}}, \bibnamefont{et~al.},
  \bibinfo{journal}{European Physical Journal C} \textbf{\bibinfo{volume}{70}},
  \bibinfo{pages}{823} (\bibinfo{year}{2010}), \eprint{1005.4568}.

\bibitem[{\citenamefont{{He} et~al.}(2016)\citenamefont{{He}, {Ma}, {Chang},
  {Zhang}, {Huang}, {Zang}, {Wu}, and {Dong}}}]{2016AcASn..57....1H}
\bibinfo{author}{\bibfnamefont{M.}~\bibnamefont{{He}}},
  \bibinfo{author}{\bibfnamefont{T.}~\bibnamefont{{Ma}}},
  \bibinfo{author}{\bibfnamefont{J.}~\bibnamefont{{Chang}}},
  \bibinfo{author}{\bibfnamefont{Y.}~\bibnamefont{{Zhang}}},
  \bibinfo{author}{\bibfnamefont{Y.~Y.} \bibnamefont{{Huang}}},
  \bibinfo{author}{\bibfnamefont{J.~J.} \bibnamefont{{Zang}}},
  \bibinfo{author}{\bibfnamefont{J.}~\bibnamefont{{Wu}}}, \bibnamefont{and}
  \bibinfo{author}{\bibfnamefont{T.~K.} \bibnamefont{{Dong}}},
  \bibinfo{journal}{Acta Astronomica Sinica} \textbf{\bibinfo{volume}{57}},
  \bibinfo{pages}{1} (\bibinfo{year}{2016}).

\bibitem[{\citenamefont{{Jiang}}(2022)}]{2022icrc.confE..82J}
\bibinfo{author}{\bibfnamefont{W.}~\bibnamefont{{Jiang}}}, in
  \emph{\bibinfo{booktitle}{37th International Cosmic Ray Conference}}
  (\bibinfo{year}{2022}), p.~\bibinfo{pages}{82}.

\bibitem[{\citenamefont{He et~al.}(2023)\citenamefont{He, Niu, Wang, Liang,
  Liu, Tian, Zhang, Zou, Liu, Zhang et~al.}}]{he2023advances}
\bibinfo{author}{\bibfnamefont{R.}~\bibnamefont{He}}, \bibnamefont{et~al.},
  \bibinfo{journal}{Nuclear Science and Techniques}
  \textbf{\bibinfo{volume}{34}}, \bibinfo{pages}{205} (\bibinfo{year}{2023}).

\bibitem[{\citenamefont{{Qiao} et~al.}(2019)\citenamefont{{Qiao}, {Peng},
  {Ambrosi}, {Asfandiyarov}, {Azzarello}, {Bernardini}, {Bertucci},
  {Bolognini}, {Cadoux}, {Caprai} et~al.}}]{2019NIMPA.935...24Q}
\bibinfo{author}{\bibfnamefont{R.}~\bibnamefont{{Qiao}}}, \bibnamefont{et~al.},
  \bibinfo{journal}{Nuclear Instruments and Methods in Physics Research A}
  \textbf{\bibinfo{volume}{935}}, \bibinfo{pages}{24} (\bibinfo{year}{2019}).

\bibitem[{\citenamefont{{Dong} et~al.}(2015)\citenamefont{{Dong}, {Zhang},
  {Qiao}, {Peng}, {Fan}, {Gong}, {Wu}, and {Wang}}}]{2015ChPhC..39k6202D}
\bibinfo{author}{\bibfnamefont{Y.-F.} \bibnamefont{{Dong}}},
  \bibinfo{author}{\bibfnamefont{F.}~\bibnamefont{{Zhang}}},
  \bibinfo{author}{\bibfnamefont{R.}~\bibnamefont{{Qiao}}},
  \bibinfo{author}{\bibfnamefont{W.-X.} \bibnamefont{{Peng}}},
  \bibinfo{author}{\bibfnamefont{R.-R.} \bibnamefont{{Fan}}},
  \bibinfo{author}{\bibfnamefont{K.}~\bibnamefont{{Gong}}},
  \bibinfo{author}{\bibfnamefont{D.}~\bibnamefont{{Wu}}}, \bibnamefont{and}
  \bibinfo{author}{\bibfnamefont{H.-Y.} \bibnamefont{{Wang}}},
  \bibinfo{journal}{Chinese Physics C} \textbf{\bibinfo{volume}{39}},
  \bibinfo{eid}{116202} (\bibinfo{year}{2015}), \eprint{1503.00415}.

\bibitem[{\citenamefont{{Zhang} et~al.}(2016)\citenamefont{{Zhang}, {Peng},
  {Gong}, {Wu}, {Dong}, {Qiao}, {Fan}, {Wang}, {Wang}, {Wu}
  et~al.}}]{2016ChPhC..40k6101Z}
\bibinfo{author}{\bibfnamefont{F.}~\bibnamefont{{Zhang}}},
  \bibnamefont{et~al.}, \bibinfo{journal}{Chinese Physics C}
  \textbf{\bibinfo{volume}{40}}, \bibinfo{eid}{116101} (\bibinfo{year}{2016}),
  \eprint{1606.05080}.

\bibitem[{\citenamefont{{Qiao} et~al.}(2018)\citenamefont{{Qiao}, {Peng},
  {Guo}, {Zhao}, {Wang}, {Gong}, {Zhang}, {Wu}, {Azzarello}, {Tykhonov}
  et~al.}}]{2018NIMPA.886...48Q}
\bibinfo{author}{\bibfnamefont{R.}~\bibnamefont{{Qiao}}}, \bibnamefont{et~al.},
  \bibinfo{journal}{Nuclear Instruments and Methods in Physics Research A}
  \textbf{\bibinfo{volume}{886}}, \bibinfo{pages}{48} (\bibinfo{year}{2018}),
  \eprint{1705.09791}.

\bibitem[{\citenamefont{{Azzarello} et~al.}(2016)\citenamefont{{Azzarello},
  {Ambrosi}, {Asfandiyarov}, {Bernardini}, {Bertucci}, {Bolognini}, {Cadoux},
  {Caprai}, {De Mitri}, {Domenjoz} et~al.}}]{2016NIMPA.831..378A}
\bibinfo{author}{\bibfnamefont{P.}~\bibnamefont{{Azzarello}}},
  \bibnamefont{et~al.}, \bibinfo{journal}{Nuclear Instruments and Methods in
  Physics Research A} \textbf{\bibinfo{volume}{831}}, \bibinfo{pages}{378}
  (\bibinfo{year}{2016}).

\bibitem[{\citenamefont{{Fr{\"u}hwirth}}(1987)}]{1987NIMPA.262..444F}
\bibinfo{author}{\bibfnamefont{R.}~\bibnamefont{{Fr{\"u}hwirth}}},
  \bibinfo{journal}{Nuclear Instruments and Methods in Physics Research A}
  \textbf{\bibinfo{volume}{262}}, \bibinfo{pages}{444} (\bibinfo{year}{1987}).

\bibitem[{\citenamefont{{Cervera-Villanueva}
  et~al.}(2002)\citenamefont{{Cervera-Villanueva}, {do Couto e Silva}, {Ellis},
  {Ferr{\`e}re}, {G{\'o}mez-Cadenas}, {Gouan{\`e}re}, {Hernando}, {Kokkonen},
  {Kuznetsov}, {Linssen} et~al.}}]{2002NIMPA.486..639C}
\bibinfo{author}{\bibfnamefont{A.}~\bibnamefont{{Cervera-Villanueva}}},
  \bibnamefont{et~al.}, \bibinfo{journal}{Nuclear Instruments and Methods in
  Physics Research A} \textbf{\bibinfo{volume}{486}}, \bibinfo{pages}{639}
  (\bibinfo{year}{2002}).

\bibitem[{\citenamefont{{Evensen}}(2003)}]{2003OcDyn..53..343E}
\bibinfo{author}{\bibfnamefont{G.}~\bibnamefont{{Evensen}}},
  \bibinfo{journal}{Ocean Dynamics} \textbf{\bibinfo{volume}{53}},
  \bibinfo{pages}{343} (\bibinfo{year}{2003}).

\bibitem[{\citenamefont{Hernando}(1998)}]{hernando1998kalman}
\bibinfo{author}{\bibfnamefont{J.~A.} \bibnamefont{Hernando}},
  \bibinfo{journal}{SCIPP} \textbf{\bibinfo{volume}{98}}, \bibinfo{pages}{18}
  (\bibinfo{year}{1998}).

\bibitem[{\citenamefont{{Lu} et~al.}(2017)\citenamefont{{Lu}, {Lei}, {Zang},
  {Chang}, and {Wu}}}]{2017ChA&A..41..455L}
\bibinfo{author}{\bibfnamefont{T.-S.} \bibnamefont{{Lu}}},
  \bibinfo{author}{\bibfnamefont{S.-J.} \bibnamefont{{Lei}}},
  \bibinfo{author}{\bibfnamefont{J.-J.} \bibnamefont{{Zang}}},
  \bibinfo{author}{\bibfnamefont{J.}~\bibnamefont{{Chang}}}, \bibnamefont{and}
  \bibinfo{author}{\bibfnamefont{J.}~\bibnamefont{{Wu}}},
  \bibinfo{journal}{Chinese Astronomy and Astrophysics}
  \textbf{\bibinfo{volume}{41}}, \bibinfo{pages}{455} (\bibinfo{year}{2017}).

\bibitem[{\citenamefont{{Yue} et~al.}(2017)\citenamefont{{Yue}, {Zang}, {Dong},
  {Li}, {Zhang}, {Zimmer}, {Jiang}, {Zhang}, and {Wei}}}]{2017NIMPA.856...11Y}
\bibinfo{author}{\bibfnamefont{C.}~\bibnamefont{{Yue}}},
  \bibinfo{author}{\bibfnamefont{J.}~\bibnamefont{{Zang}}},
  \bibinfo{author}{\bibfnamefont{T.}~\bibnamefont{{Dong}}},
  \bibinfo{author}{\bibfnamefont{X.}~\bibnamefont{{Li}}},
  \bibinfo{author}{\bibfnamefont{Z.}~\bibnamefont{{Zhang}}},
  \bibinfo{author}{\bibfnamefont{S.}~\bibnamefont{{Zimmer}}},
  \bibinfo{author}{\bibfnamefont{W.}~\bibnamefont{{Jiang}}},
  \bibinfo{author}{\bibfnamefont{Y.}~\bibnamefont{{Zhang}}}, \bibnamefont{and}
  \bibinfo{author}{\bibfnamefont{D.}~\bibnamefont{{Wei}}},
  \bibinfo{journal}{Nuclear Instruments and Methods in Physics Research A}
  \textbf{\bibinfo{volume}{856}}, \bibinfo{pages}{11} (\bibinfo{year}{2017}),
  \eprint{1703.02821}.

\bibitem[{\citenamefont{{Baldini}}(2014)}]{2014arXiv1407.7631B}
\bibinfo{author}{\bibfnamefont{L.}~\bibnamefont{{Baldini}}},
  \bibinfo{journal}{arXiv e-prints} \bibinfo{eid}{arXiv:1407.7631}
  (\bibinfo{year}{2014}), \eprint{1407.7631}.

\bibitem[{\citenamefont{{Caputo} et~al.}(2017)\citenamefont{{Caputo}, {Kislat},
  {Racusin}, and {Amego Team}}}]{2017ICRC...35..783C}
\bibinfo{author}{\bibfnamefont{R.}~\bibnamefont{{Caputo}}},
  \bibinfo{author}{\bibfnamefont{F.}~\bibnamefont{{Kislat}}},
  \bibinfo{author}{\bibfnamefont{J.}~\bibnamefont{{Racusin}}},
  \bibnamefont{and} \bibinfo{author}{\bibnamefont{{Amego Team}}}, in
  \emph{\bibinfo{booktitle}{35th International Cosmic Ray Conference
  (ICRC2017)}} (\bibinfo{year}{2017}), vol. \bibinfo{volume}{301} of
  \emph{\bibinfo{series}{International Cosmic Ray Conference}}, p.
  \bibinfo{pages}{783}.

\bibitem[{\citenamefont{{Fermi-LAT Collaboration}}()}]{Fermi_webpage}
\bibinfo{author}{\bibnamefont{{Fermi-LAT Collaboration}}},
  \emph{\bibinfo{title}{{Fermi LAT Performance}}},
  \bibinfo{howpublished}{\url{https://www.slac.stanford.edu/exp/glast/groups/c%
anda/lat_Performance.htm.}}

\bibitem[{\citenamefont{{Ackermann}
  et~al.}(2013{\natexlab{b}})\citenamefont{{Ackermann}, {Ajello}, {Albert},
  {Allafort}, {Baldini}, {Barbiellini}, {Bastieri}, {Bechtol}, {Bellazzini},
  {Bissaldi} et~al.}}]{2013PhRvD..88h2002A}
\bibinfo{author}{\bibfnamefont{M.}~\bibnamefont{{Ackermann}}},
  \bibnamefont{et~al.}, \bibinfo{journal}{Physical Review D}
  \textbf{\bibinfo{volume}{88}}, \bibinfo{eid}{082002}
  (\bibinfo{year}{2013}{\natexlab{b}}), \eprint{1305.5597}.

\bibitem[{\citenamefont{{Liang}}(2022)}]{2022SciBu..67..679L}
\bibinfo{author}{\bibfnamefont{Y.-F.} \bibnamefont{{Liang}}},
  \bibinfo{journal}{Science Bulletin} \textbf{\bibinfo{volume}{67}},
  \bibinfo{pages}{679} (\bibinfo{year}{2022}), \eprint{2112.08860}.

\bibitem[{\citenamefont{{Chen} and {Zhou}}(2013)}]{2013JCAP...04..017C}
\bibinfo{author}{\bibfnamefont{J.}~\bibnamefont{{Chen}}} \bibnamefont{and}
  \bibinfo{author}{\bibfnamefont{Y.-F.} \bibnamefont{{Zhou}}},
  \bibinfo{journal}{Journal of Cosmology and Astroparticle Physics}
  \textbf{\bibinfo{volume}{2013}}, \bibinfo{eid}{017} (\bibinfo{year}{2013}),
  \eprint{1301.5778}.

\bibitem[{\citenamefont{{Zhang} et~al.}(2015)\citenamefont{{Zhang}, {Zhang},
  {Dong}, {Wen}, {Feng}, {Wang}, {Wei}, {Wang}, {Xu}, and
  {Liu}}}]{2015NIMPA.780...21Z}
\bibinfo{author}{\bibfnamefont{Z.}~\bibnamefont{{Zhang}}},
  \bibinfo{author}{\bibfnamefont{Y.}~\bibnamefont{{Zhang}}},
  \bibinfo{author}{\bibfnamefont{J.}~\bibnamefont{{Dong}}},
  \bibinfo{author}{\bibfnamefont{S.}~\bibnamefont{{Wen}}},
  \bibinfo{author}{\bibfnamefont{C.}~\bibnamefont{{Feng}}},
  \bibinfo{author}{\bibfnamefont{C.}~\bibnamefont{{Wang}}},
  \bibinfo{author}{\bibfnamefont{Y.}~\bibnamefont{{Wei}}},
  \bibinfo{author}{\bibfnamefont{X.}~\bibnamefont{{Wang}}},
  \bibinfo{author}{\bibfnamefont{Z.}~\bibnamefont{{Xu}}}, \bibnamefont{and}
  \bibinfo{author}{\bibfnamefont{S.}~\bibnamefont{{Liu}}},
  \bibinfo{journal}{Nuclear Instruments and Methods in Physics Research A}
  \textbf{\bibinfo{volume}{780}}, \bibinfo{pages}{21} (\bibinfo{year}{2015}).

\bibitem[{\citenamefont{{Feng} et~al.}(2015)\citenamefont{{Feng}, {Zhang},
  {Zhang}, {Gao}, {Yang}, {Zhang}, {Zhang}, {Liu}, and
  {An}}}]{2015ITNS...62.3117F}
\bibinfo{author}{\bibfnamefont{C.}~\bibnamefont{{Feng}}},
  \bibinfo{author}{\bibfnamefont{D.}~\bibnamefont{{Zhang}}},
  \bibinfo{author}{\bibfnamefont{J.}~\bibnamefont{{Zhang}}},
  \bibinfo{author}{\bibfnamefont{S.}~\bibnamefont{{Gao}}},
  \bibinfo{author}{\bibfnamefont{D.}~\bibnamefont{{Yang}}},
  \bibinfo{author}{\bibfnamefont{Y.}~\bibnamefont{{Zhang}}},
  \bibinfo{author}{\bibfnamefont{Z.}~\bibnamefont{{Zhang}}},
  \bibinfo{author}{\bibfnamefont{S.}~\bibnamefont{{Liu}}}, \bibnamefont{and}
  \bibinfo{author}{\bibfnamefont{Q.}~\bibnamefont{{An}}},
  \bibinfo{journal}{IEEE Transactions on Nuclear Science}
  \textbf{\bibinfo{volume}{62}}, \bibinfo{pages}{3117} (\bibinfo{year}{2015}).

\bibitem[{\citenamefont{{Zhao} et~al.}(2022)\citenamefont{{Zhao}, {Dai}, {Liu},
  {Wei}, {Zhang}, {Zhang}, {Wu}, {Wang}, {Wang}, {Xu}
  et~al.}}]{2022NIMPA102966453Z}
\bibinfo{author}{\bibfnamefont{C.}~\bibnamefont{{Zhao}}}, \bibnamefont{et~al.},
  \bibinfo{journal}{Nuclear Instruments and Methods in Physics Research A}
  \textbf{\bibinfo{volume}{1029}}, \bibinfo{eid}{166453}
  (\bibinfo{year}{2022}).

\bibitem[{\citenamefont{{Wei} et~al.}(2019)\citenamefont{{Wei}, {Zhang},
  {Zhang}, {Wu}, {Wen}, {Dai}, {Liu}, {Wang}, {Xu}, {Huang}
  et~al.}}]{2019NIMPA.922..177W}
\bibinfo{author}{\bibfnamefont{Y.}~\bibnamefont{{Wei}}}, \bibnamefont{et~al.},
  \bibinfo{journal}{Nuclear Instruments and Methods in Physics Research A}
  \textbf{\bibinfo{volume}{922}}, \bibinfo{pages}{177} (\bibinfo{year}{2019}).

\bibitem[{\citenamefont{{DAMPE Collaboration} et~al.}(2017)\citenamefont{{DAMPE
  Collaboration}, {Ambrosi}, {An}, {Asfandiyarov}, {Azzarello}, {Bernardini},
  {Bertucci}, {Cai}, {Chang}, {Chen} et~al.}}]{2017Natur.552...63D}
\bibinfo{author}{\bibnamefont{{DAMPE Collaboration}}}, \bibnamefont{et~al.},
  \bibinfo{journal}{Nature} \textbf{\bibinfo{volume}{552}}, \bibinfo{pages}{63}
  (\bibinfo{year}{2017}), \eprint{1711.10981}.

\bibitem[{\citenamefont{{Jiang} et~al.}(2020)\citenamefont{{Jiang}, {Yue},
  {Cui}, {Li}, {Yuan}, {Alemanno}, {Bernardini}, {Catanzani}, {Chen}, {De
  Mitri} et~al.}}]{2020ChPhL..37k9601J}
\bibinfo{author}{\bibfnamefont{W.}~\bibnamefont{{Jiang}}},
  \bibnamefont{et~al.}, \bibinfo{journal}{Chinese Physics Letters}
  \textbf{\bibinfo{volume}{37}}, \bibinfo{eid}{119601} (\bibinfo{year}{2020}),
  \eprint{2009.13036}.

\bibitem[{\citenamefont{{Xu} et~al.}(2022)\citenamefont{{Xu}, {Li}, {Cui},
  {Yue}, {Jiang}, {Li}, and {Yuan}}}]{2022Univ....8..570X}
\bibinfo{author}{\bibfnamefont{Z.}~\bibnamefont{{Xu}}},
  \bibinfo{author}{\bibfnamefont{X.}~\bibnamefont{{Li}}},
  \bibinfo{author}{\bibfnamefont{M.}~\bibnamefont{{Cui}}},
  \bibinfo{author}{\bibfnamefont{C.}~\bibnamefont{{Yue}}},
  \bibinfo{author}{\bibfnamefont{W.}~\bibnamefont{{Jiang}}},
  \bibinfo{author}{\bibfnamefont{W.}~\bibnamefont{{Li}}}, \bibnamefont{and}
  \bibinfo{author}{\bibfnamefont{Q.}~\bibnamefont{{Yuan}}},
  \bibinfo{journal}{Universe} \textbf{\bibinfo{volume}{8}}, \bibinfo{eid}{570}
  (\bibinfo{year}{2022}), \eprint{2212.01843}.

\bibitem[{\citenamefont{{Ajello} et~al.}(2019)\citenamefont{{Ajello},
  {Arimoto}, {Axelsson}, {Baldini}, {Barbiellini}, {Bastieri}, {Bellazzini},
  {Bhat}, {Bissaldi}, {Blandford} et~al.}}]{2019ApJ...878...52A}
\bibinfo{author}{\bibfnamefont{M.}~\bibnamefont{{Ajello}}},
  \bibnamefont{et~al.}, \bibinfo{journal}{\apj} \textbf{\bibinfo{volume}{878}},
  \bibinfo{eid}{52} (\bibinfo{year}{2019}), \eprint{1906.11403}.

\bibitem[{\citenamefont{{Kang} et~al.}(2020)\citenamefont{{Kang}, {Qiao},
  {Yao}, {Guo}, {Hu}, and {Yao}}}]{2020ApJ...900...67K}
\bibinfo{author}{\bibfnamefont{M.-M.} \bibnamefont{{Kang}}},
  \bibinfo{author}{\bibfnamefont{B.-Q.} \bibnamefont{{Qiao}}},
  \bibinfo{author}{\bibfnamefont{Y.-H.} \bibnamefont{{Yao}}},
  \bibinfo{author}{\bibfnamefont{Y.-Q.} \bibnamefont{{Guo}}},
  \bibinfo{author}{\bibfnamefont{H.-B.} \bibnamefont{{Hu}}}, \bibnamefont{and}
  \bibinfo{author}{\bibfnamefont{Z.-G.} \bibnamefont{{Yao}}},
  \bibinfo{journal}{\apj} \textbf{\bibinfo{volume}{900}}, \bibinfo{eid}{67}
  (\bibinfo{year}{2020}).

\bibitem[{\citenamefont{{Xin} et~al.}(2021)\citenamefont{{Xin}, {Yao}, {Qian},
  {Liu}, {Gao}, {Luo-Bu}, {Feng}, {Gou}, {Hu}, {Li}
  et~al.}}]{2021ApJ...923..112X}
\bibinfo{author}{\bibfnamefont{G.-G.} \bibnamefont{{Xin}}},
  \bibnamefont{et~al.}, \bibinfo{journal}{\apj} \textbf{\bibinfo{volume}{923}},
  \bibinfo{eid}{112} (\bibinfo{year}{2021}), \eprint{2103.04381}.

\bibitem[{\citenamefont{{Yao} et~al.}(2023)\citenamefont{{Yao}, {Wang}, {Chen},
  {Chen}, {Feng}, {Gao}, {Gou}, {Guo}, {Hu}, {Kang}
  et~al.}}]{2023ApJ...958...87Y}
\bibinfo{author}{\bibfnamefont{Y.-H.} \bibnamefont{{Yao}}},
  \bibnamefont{et~al.}, \bibinfo{journal}{\apj} \textbf{\bibinfo{volume}{958}},
  \bibinfo{eid}{87} (\bibinfo{year}{2023}).

\end{thebibliography}

\end{document}